\documentclass[aps,twocolumn,prb,amsmath,amssymb]{revtex4-1}

\usepackage{graphicx}
\usepackage{color}
\usepackage{natbib}
\usepackage{hyperref}
\usepackage{notes2bib}
\usepackage{mathbbol}
\usepackage{float}
\usepackage{ulem}

\def\a{\alpha}

\def\dg{{\;\dagger}}

\def\vep{\varepsilon}

\def\m{\mathcal}
\def\|{\ll}

\def\ll{\langle\langle}

\def\doubleunderline#1{\underline{\underline{#1}}}

\begin{document}

\parindent=0pt

\title{Dynamical charge susceptibility in nonequilibrium double quantum dots}
\author{A. Cr\'epieux$^{1}$}
\author{M. Lavagna$^{2,3}$}
\affiliation{$^1$ Aix Marseille Univ, Universit\'e de Toulon, CNRS, CPT, Marseille, France}
\affiliation{$^2$ Univ. Grenoble Alpes, CEA, IRIG, PHELIQS, F-38000 Grenoble, France}
\affiliation{$^3$ Centre National de la Recherche Scientifique, CNRS, 38042 Grenoble, France}
%\date{\today}

\begin{abstract}
Double quantum dots are one of the promising two-state quantum systems for realizing qubits. In the quest of successfully manipulating and reading information in qubit systems, it is of prime interest to control the charge response of the system to a gate voltage, as filled in by the dynamical charge susceptibility. We theoretically study this quantity for a nonequilibrium double quantum~dot by using the functional integral approach and derive its general analytical expression. One highlights the existence of two lines of maxima as a function of the dot level energies, each of them being split under the action of a bias voltage. In the low frequency limit, we derive the capacitance and the charge relaxation resistance of the equivalent quantum RC-circuit with a notable difference in the range of variation for $R$ depending on whether the system is connected in series or in parallel. By~incorporating an additional triplet state in order to describe the situation of a double quantum dot with spin, we obtain results for the resonator phase response which are in qualitative agreement with recent experimental observations in spin qubit systems. These results show the wealth of information brought by the knowledge of dynamical charge susceptibility in double quantum dots with potential applications for spin qubits.
\end{abstract}

\maketitle

%%%%%%%%%%%%%%%%%%%%%%%%%%%%%%%%%%%%%%%%%%%%%%%%%%%%%%%%%%%%%%%%%%
%																 %
%																 %
%		INTRODUCTION											 %
%																 %
%																 %
%%%%%%%%%%%%%%%%%%%%%%%%%%%%%%%%%%%%%%%%%%%%%%%%%%%%%%%%%%%%%%%%%%

\section{Introduction}

In double quantum dots (DQDs), the knowledge of the dynamical charge susceptibility (DCS), which measures the ability of a system to adapt its electronic charge to an ac gate-voltage, is of fundamental interest in the general context of circuit quantum electrodynamics with gated GaAs, silicon, and germanium semiconductor quantum dots. This field has become all the more important because of its expected implications in manipulation, control and readout of spin qubits\cite{Petersson2012}. There certainly are some theoretical works on charge susceptibility in~DQD but they are mainly restricted to the study at zero frequency\cite{Galpin2006,Talbo2018,Lavagna2020} and in the low frequency regime with the determination of mesoscopic admittance\cite{Cottet2011} and quantum capacitance\cite{Mizuta2017}, or to calculations performed at the lowest order in dot-lead coupling amplitude\cite{Bruhat2018}. Electrical transport experiments in~DQD systems are, however, restricted neither to the weak dot-lead coupling regime nor to the measurement at low frequencies. On the contrary, in the last ten years one has witnessed a considerable experimental effort\cite{Schroer2012,Chorley2012,Colless2013,Viennot2014,Mi2018,Harvey2018,Scarlino2019,Lundberg2020,Ezzouch2021} with measurements performed by using either on-chip superconducting resonant detectors\cite{Zheng2019,Jong2021,Scarlino2021} or dispersive probed microwave spectroscopy via reflectometry techniques\cite{Crippa2019,Bugu2021,Filmer2022}, all of them made in the high frequency regime. To accompany this growing experimental development, it becomes of primary importance to progress on the theoretical level in order to investigate circuit quantum electrodynamics at high frequency in nanoscale systems. Indeed, the interpretation of these experiments requires the precise knowledge of the DCS at any frequency and temperature range, in both equilibrium and out-of-equilibrium DQDs. This article is precisely devoted to this theoretical issue. It is organized as follows: in Sec.~II we present the functional integral approach used to solve this problem and give the expression for the dynamical charge susceptibility, in Sec.~III we give the results for both serial and parallel DQDs, in Sec.~IV we focus on the characterization of the equivalent quantum RC-circuit to the DQD system. Finally, in Sec.~V, we study the reflection phase of the system considered as a resonator embedded in an electromagnetic environment, shedding new light on recent measurements performed in microwave reflectometry experiments in spin qubit systems. We conclude in Sec.~VI.

%%%%%%%%%%%%%%%%%%%%%%%%%%%%%%%%%%%%%%%%%%%%%%%%%%%%%%%%%%%%%%%%%%
%																 %
%																 %
%		MODEL  											 %
%																 %
%																 %
%%%%%%%%%%%%%%%%%%%%%%%%%%%%%%%%%%%%%%%%%%%%%%%%%%%%%%%%%%%%%%%%%%

\section{Model}

\subsection{Functional integral approach}

\begin{figure}
\begin{center}
\includegraphics[width=8cm]{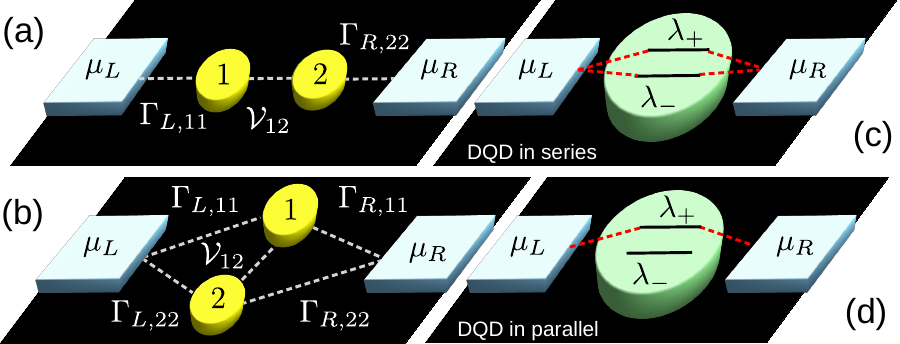}
\caption{Schematic view of the DQD in the (a),~(c) serial and (b),~(d) parallel geometries. At $\varepsilon_1=\varepsilon_2$ and for symmetrical couplings to the leads, the bonding state~$\lambda_-$ is disconnected from the leads in the parallel geometry.}\label{figure0}
\end{center}
\end{figure}

Let us consider a DQD connected to two leads described by the Hamiltonian $\m{\widehat H}=\m{\widehat H}_\text{leads}+\m{\widehat H}_\text{dots}+\m{\widehat H}_\text{hop}$, with $\m{\widehat H}_\text{leads} = \sum_{\substack{\a =L,R ; k \in \a}} \vep_{\a k}\widehat c_{\a k}^\dg \widehat c_{\a k}$, $\m{\widehat H}_\text{dots} = \sum_{i=1,2}\vep_{i}\widehat d_{i}^\dg \widehat d_{i}+\m{V}_{12}\widehat d_{2}^\dg \widehat d_{1} +\m{V}_{21}\widehat d_{1}^\dg \widehat d_{2} $, and $\m{\widehat H}_\text{hop}=\sum_{\substack{\a =L,R ; k \in \a}}\sum_{i=1,2}V_{i,\a k}\widehat c_{\a k}^\dg \widehat d_{i} +H.c.$, where $\vep_{i}$ is the energy level of the dot $i$, $\m{V}_{12}$ is the interdot coupling, $\vep_{\a k}$ is the energy of one electron with momentum $k$ in the lead~$\alpha$, and $V_{i,\a k}$ is the hopping energy between the dot~$i$ and the lead~$\alpha$ for momentum $k$. The very general form considered in the expression for $\mathcal{\widehat H}$ allows one to describe the situations where the two dots are connected in series as well as in parallel (see Fig.~\ref{figure0}). We use the functional integral approach to derive the expression for the DCS. The partition function of the system writes
\begin{eqnarray}\label{lagrangian}
 \mathcal{Z}=\int \prod_{i=1,2}\text{d}d_{i}^\dg \text{d}d_{i}\prod_{\substack{\a =L,R\\k \in \a}}\text{d}c_{\a k}^\dg \text{d}c_{\a k}e^{-\int_0^{\beta}d\tau\mathcal{L}(\tau)},
\end{eqnarray}
where $\mathcal{L}(\tau)=\sum_{i=1,2}\widehat d_{i}^\dg\partial_\tau \widehat d_{i}+\sum_{\substack{\a =L,R ; k \in \a}}\widehat c_{\alpha k}^\dg\partial_\tau \widehat c_{\alpha k}-\widehat {\m H}$ is the Lagrangian, $d_{i}^{\,(\dag)}$ and $c_{\a k}^{\,(\dag)}$, the Grassmann variables associated with the operators $\widehat d_{i}^{\,(\dag)}$ and $\widehat c_{\a k}^{\,(\dag)}$. By integrating over the Grassmann variables $\widetilde c_{\alpha k}=c_{\alpha k}-\sum_{j=1,2}V_{j,\alpha k}(\partial_\tau+\varepsilon_{\alpha k})^{-1}d_j$\cite{SM}, one gets $ \mathcal{Z}=\int \prod_{i=1,2}\text{d}d_{i}^\dg \text{d}d_{i}e^{-\int_0^{\beta}d\tau\mathcal{L}_\text{eff}(\tau)}$, where $\mathcal{L}_\text{eff}(\tau)$ is an effective Lagrangian defined as
\begin{eqnarray}\label{effective_lagrangian}
 &&\mathcal{L}_\text{eff}(\tau)=
\left(
\begin{array}{c}
d_1^\dg\;\;d_2^\dg
\end{array}
\right)\nonumber\\
&&\times\left(
\begin{array}{cc}
\partial_\tau+\varepsilon_1+{\mathbb{\Sigma}}_{11}(\tau)& \widetilde{\mathbb{\Sigma}}_{12}(\tau)\\
\widetilde{\mathbb{\Sigma}}_{21}(\tau)&\partial_\tau+\varepsilon_2+{\mathbb{\Sigma}}_{22}(\tau)
\end{array}
\right)
\left(
\begin{array}{c}
d_1\\
d_2
\end{array}
\right),
\end{eqnarray}
where ${\mathbb{\Sigma}}_{ij}(\tau)=\sum_{\a =L,R}\sum_{k \in \a}V_{i,\alpha k}^*g_{\alpha k}(\tau) V_{j,\alpha k}$, $\widetilde {\mathbb{\Sigma}}_{i\overline{i}}(\varepsilon)={\mathbb{\Sigma}}_{i\overline{i}}(\varepsilon)+{\m V}_{i\overline{i}}^*$, within the notation $\overline{1}=2$ and $\overline{2}=1$, and $g_{\a k}(\tau)=-(\partial_\tau+\varepsilon_{\alpha k})^{-1}$ is the Green function in the lead~$\alpha$ for momentum $k$. In the wide-flat-band limit for electrons in the leads and when $V_{i,\alpha k}$ is assumed to be $k$-independent, one has $\mathbb{\Sigma}^r_{ij}(\varepsilon)=-i\Gamma_{ij}/2$ and $\Gamma_{ij}=\sum_{\alpha=L,R}\Gamma_{\alpha,ij}$, where $\Gamma_{\alpha,ij}=2\pi V^*_{i,\alpha k}V_{j,\alpha k}\rho_\alpha$. The density of states $\rho_\alpha$ in the lead~$\alpha$ is assumed to be equal to $W^{-1}$, where $W$ is the energy band width taken as the energy unit in the rest of the article. From Eq.~(\ref{effective_lagrangian}) one extracts  an effective Hamiltonian given by
\begin{eqnarray}
\mathcal{H}_\text{eff}=\left(
\begin{array}{ccc}
\varepsilon_1-i\Gamma_{11}/2& & \mathcal{V}_{12}^{\,*}-i\Gamma_{12}/2\\
\mathcal{V}_{21}^{\,*}-i\Gamma_{21}/2& &\varepsilon_2-i\Gamma_{22}/2
\end{array}
\right)~,
\end{eqnarray}
which can be diagonalized leading to the eigenenergies
\begin{eqnarray}\label{def_lambda}
\lambda_\pm= \frac{1}{2}\left(\varepsilon_1-i\frac{\Gamma_{11}}{2}+\varepsilon_2-i\frac{\Gamma_{22}}{2}\pm \Delta\right),
\end{eqnarray}
with
$\Delta^2=(\varepsilon_1-i\Gamma_{11}/2-\varepsilon_2+i\Gamma_{22}/2)^2
 +4( \mathcal{V}_{12}^{\,*}-i\Gamma_{12}/2) (\mathcal{V}_{21}^{\,*}-i\Gamma_{21}/2)$. The corresponding eigenstates are the bonding and anti-bonding states of the DQD. They have a finite relaxation rate related to the imaginary parts of $\lambda_\pm$, resulting from energy dissipation through connections to leads\cite{Rotter2009}.

%%%%%%%%%%%%%%%%%%%%%%%%%%%%%%%%%%%%%%%%%%%%%%%%%%%%%%%%%%%%%%%%%%
%																 %
%																 %
%		DYNAMICAL CHARGE SUSCEPTIBILITY    						 %
%																 %
%																 %
%%%%%%%%%%%%%%%%%%%%%%%%%%%%%%%%%%%%%%%%%%%%%%%%%%%%%%%%%%%%%%%%%%

\subsection{Dynamical charge susceptibility}

For a DQD the charge susceptibility is $\mathcal{X}_c(t,t')=\sum_{i,j=1,2}\alpha_i\alpha_j\mathcal{X}_{ij}(t,t')$, where $\alpha_{1,2}$ are the lever-arm coefficients measuring the asymmetry of the capacitive couplings of each of the two dots to the gate voltage\cite{Lavagna2020}. In the linear response theory it is related  to a correlation function through a Kubo-type formula $\mathcal{X}_{ij}(t,t')=i\Theta(t-t')\langle[\Delta \widehat N_i(t),\Delta \widehat N_j(t')]\rangle$ with  $\widehat N_i=\widehat d_i^{\;\dag}\widehat d_i$. Decomposing the correlation function in the eigenstate basis and taking the Fourier transform, for the DCS one gets: $\mathcal{X}_c(\omega)=\sum_{s_1,s_2=\pm}\mathcal{C}_{s_1s_2}\mathcal{X}_{s_1s_2}(\omega)$, where $\mathcal{C}_{s_1s_2}$ are coherence factors defined in Ref.~\onlinecite{SM}. $ \mathcal{X}_{s_1s_2}(\omega)$ can be expressed as a function of the Green functions in the dots following
\begin{eqnarray}\label{chi_DQD_green}
\mathcal{X}_{s_1s_2}(\omega)=i\int_{-\infty}^\infty \frac{d\varepsilon}{2\pi} &&\Big[\mathcal{G}^<_{s_1s_2}(\varepsilon)\mathcal{G}^a_{s_2}(\varepsilon-\hbar\omega)\nonumber\\
&&+\mathcal{G}^r_{s_1}(\varepsilon+\hbar\omega)\mathcal{G}^<_{s_1s_2}(\varepsilon)\Big]~,
\end{eqnarray}
where $ \doubleunderline{\mathcal{G}}^{r,a,<}$ are retarded, advanced, and nonequilibrium Green functions in the eigenstate basis. $ \doubleunderline{\mathcal{G}}^{r,a}$ are diagonal matrices of elements $\mathcal{G}^r_{\pm}(\varepsilon)=(\varepsilon-\lambda_\pm)^{-1}$ and $\mathcal{G}^a_{\pm}(\varepsilon)=(\varepsilon-\lambda_\pm^*)^{-1}$. In the steady state, one has $\doubleunderline{\mathcal{G}}^<(\varepsilon)=\doubleunderline{\mathcal{G}}^r(\varepsilon)\doubleunderline U^{-1}\doubleunderline{\mathbb{\Sigma}}^<(\varepsilon)\,\doubleunderline U^{\,\text{ }}\doubleunderline{\mathcal{G}}^a(\varepsilon)$,
where $\doubleunderline{U}$ is the transition matrix from the initial state basis to the eigenstate basis, $\doubleunderline{\mathbb{\Sigma}}^<(\varepsilon)=i\sum_{\alpha=L,R}f_\alpha(\varepsilon)\doubleunderline\Gamma_{\,\alpha}$, the nonequilibrium self-energy, and $f_{\alpha}(\varepsilon)=[1+\exp((\varepsilon-\mu_{\alpha})/k_BT)]^{-1}$, the Fermi-Dirac distribution in the lead $\alpha$ of chemical potential $\mu_{\alpha}$ and temperature $T$. We have thus established a general expression for the DCS of a nonequilibrium DQD, valid whatever its geometry is.

%It is expressed as a linear combination of the correlation functions $\mathcal{X}_{s_1s_2}(\omega)$ along the bonding and anti-bonding state basis. In the rest of this Letter we present the results for $\mathcal{X}_c(\omega)$ first in the case in series, and secondly in parallel. Next we discuss the results in terms of capacitance and charge relaxation resistance of the equivalent quantum RC-circuit. Finally a comparison with available experiments performed at finite frequency in spin qubits is made.

%%%%%%%%%%%%%%%%%%%%%%%%%%%%%%%%%%%%%%%%%%%%%%%%%%%%%%%%%%%%%%%%%%
%																 %
%																 %
%		DISCUSSION ON 3D GRAPHS									 %
%																 %
%																 %
%%%%%%%%%%%%%%%%%%%%%%%%%%%%%%%%%%%%%%%%%%%%%%%%%%%%%%%%%%%%%%%%%%

\section{Results and discussion}

\subsection{DQD in series}

\begin{figure}[t]
\begin{center}
\includegraphics[width=8.5cm]{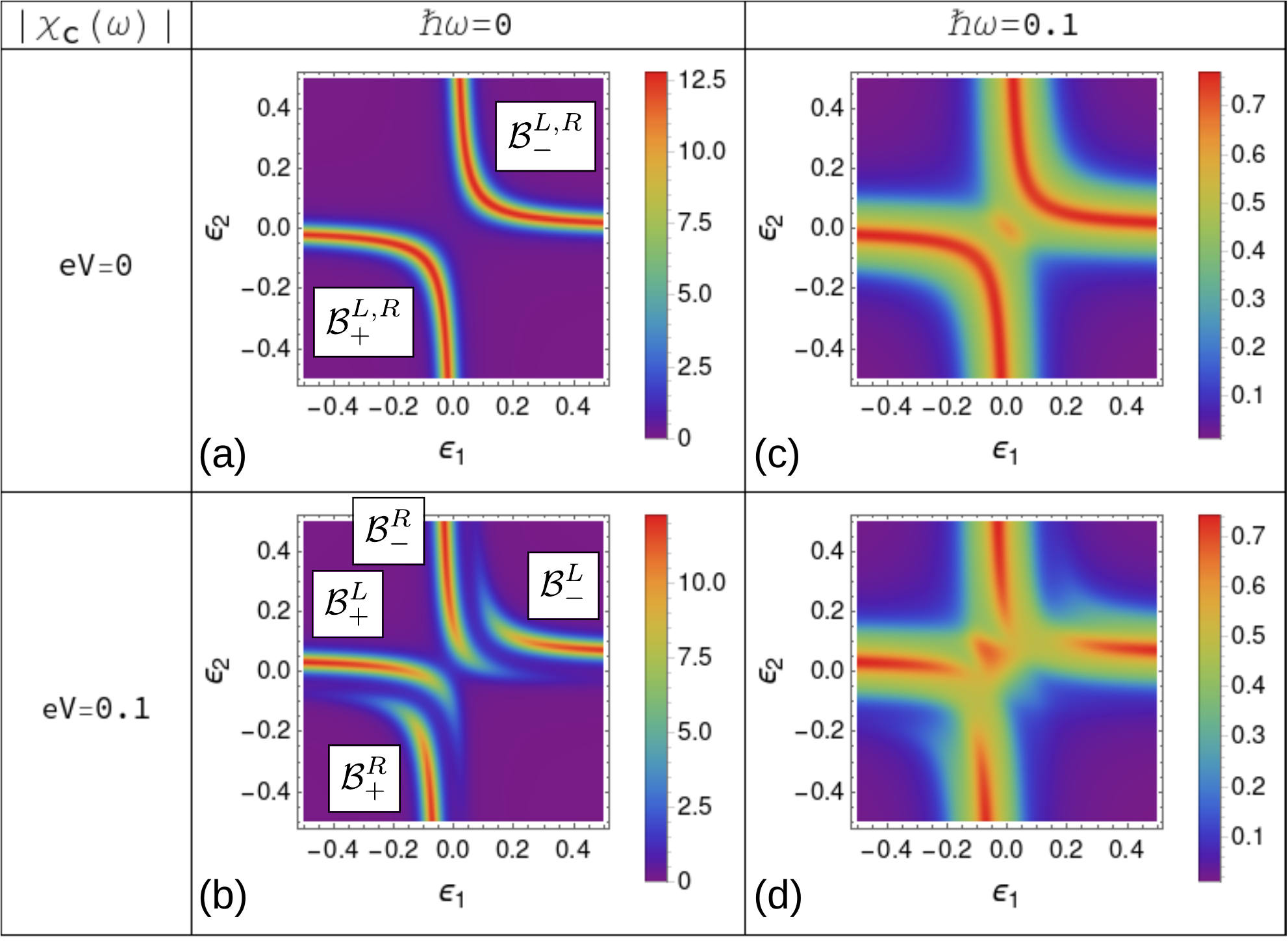}
\caption{Color-scale plots of $|\mathcal{X}_c(\omega)|$  as a function of~$\varepsilon_1$ and~$\varepsilon_2$ for a serial DQD at various values of $\omega$ and~$V$, with $\mu_L=eV/2$ and $\mu_R=-eV/2$, and $\Gamma=0.01$, $\mathcal{V}_{12}=0.1$, $k_BT=0.01$. }\label{figure1}
\end{center}
\end{figure}

The results obtained for a DQD symmetrically connected in series are shown in Fig.~\ref{figure1}. The variation of the absolute value of the DCS, $|\mathcal{X}_c(\omega)|$, is plotted in the form of color-scale plots as a function of $\varepsilon_1$ and $\varepsilon_2$. Figures~\ref{figure1}(a) and~(b) show the presence of peaks in $|\mathcal{X}_c(0)|$ along four branches denoted as $\mathcal{B}_-^{L}$, $\mathcal{B}_-^{R}$, $\mathcal{B}_+^{L}$ and~$\mathcal{B}_+^{R}$ corresponding to the alignment of the bonding and anti-bonding state energies with the chemical potential in the leads, occurring when $\text{Re}\{\lambda_\pm\}=\mu_{L,R}$\cite{Lavagna2020}. According to Eq.~(\ref{def_lambda}), one has
\begin{eqnarray}\label{lambda_series}
\lambda_\pm= \frac{1}{2}\left(\varepsilon_1+\varepsilon_2-i\Gamma\pm\sqrt{(\varepsilon_1-\varepsilon_2)^2+4|\mathcal{V}_{12}|^2}\right),
\end{eqnarray}
making use of $\Gamma_{L,11}=\Gamma_{R,22}\equiv \Gamma$, $\Gamma_{L,22}=\Gamma_{R,11}=0$, and $\Gamma_{L,12}=\Gamma_{R,12}=0$, and assuming $\alpha_1=\alpha_2=1$ and $\mathcal{V}_{12}=\mathcal{V}_{21}$. At zero voltage, $\mathcal{B}_-^{L}$ and $\mathcal{B}_-^{R}$ coincide (the same for $\mathcal{B}_+^{L}$ and $\mathcal{B}_+^{R}$). $\mathcal{B}_-^{L,R}$ and $\mathcal{B}_+^{L,R}$ are two branches of a hyperbole of equations $\varepsilon_1\varepsilon_2=|\mathcal{V}_{12}|^2$, separated by a minimal distance along the diagonal~$\mathcal{D}$ of equation $\varepsilon_1=\varepsilon_2$, taking the value of $2|\mathcal{V}_{12}|$. From Eq.~(\ref{lambda_series}), the imaginary part of $\lambda_\pm$ are both equal to $-\Gamma/2$ and independent of $\varepsilon_1$ and~$\varepsilon_2$. As a result, the width of the charge susceptibility peak arcs is uniform along the branches $\mathcal{B}_{\pm}^{L,R}$, as observed in Fig.~\ref{figure1}(a). At finite voltage, Fig.~\ref{figure1}(b) shows the splitting of the peak arcs into two branches $\mathcal{B}_-^{L}$ and $\mathcal{B}_-^{R}$, respectively $\mathcal{B}_+^{L}$ and~$\mathcal{B}_+^{R}$, with the reduction of the intensity along half of the arcs due to the fact that for a serial DQD, only the dot~1 is connected to lead~$L$ and the dot~2 to lead~$R$. Thus, only the horizontal end of $\mathcal{B}_\pm^{L}$ or the vertical end of $\mathcal{B}_\pm^{R}$ lead to a significant value of $|\mathcal{X}_c(0)|$. At finite frequency, one observes a broadening of the peak arcs located around the $\mathcal{B}_\pm^{L,R}$ branches, together with the formation of an additional central peak at $\varepsilon_1=\varepsilon_2=0$ (see Fig.~\ref{figure1}(c) and~(d)). An exact expression for~$\mathcal{X}_c(\omega)$ is derived from Eq.~(\ref{chi_DQD_green}) at $T=0$ when $\varepsilon_1=\varepsilon_2$ and $\mathcal{V}_{21}=\mathcal{V}_{12}$ for a serial DQD\cite{SM}. It writes $\mathcal{X}_c(\omega)=\sum_\pm\sum_{\alpha=L,R}\mathcal{X}_{\pm,\alpha}(\omega)$ with
\begin{eqnarray}\label{Xc_series}
 \mathcal{X}_{\pm,\alpha}(\omega)=-\frac{\Gamma}{2h\omega(\hbar\omega+i\Gamma)}
\ln\Big(1-\frac{h\omega(\hbar\omega+i\Gamma)}{\Gamma}A_\pm(\mu_\alpha)\Big),\nonumber\\
\end{eqnarray}
 where $A_\pm(\varepsilon)=-\text{Im}\{\mathcal{G}^r_{\pm}(\varepsilon)\}/\pi$ are the spectral function contributions from the anti-bonding and bonding states.

 \subsection{DQD in parallel}

\begin{figure}[t]
\begin{center}
\includegraphics[width=8.5cm]{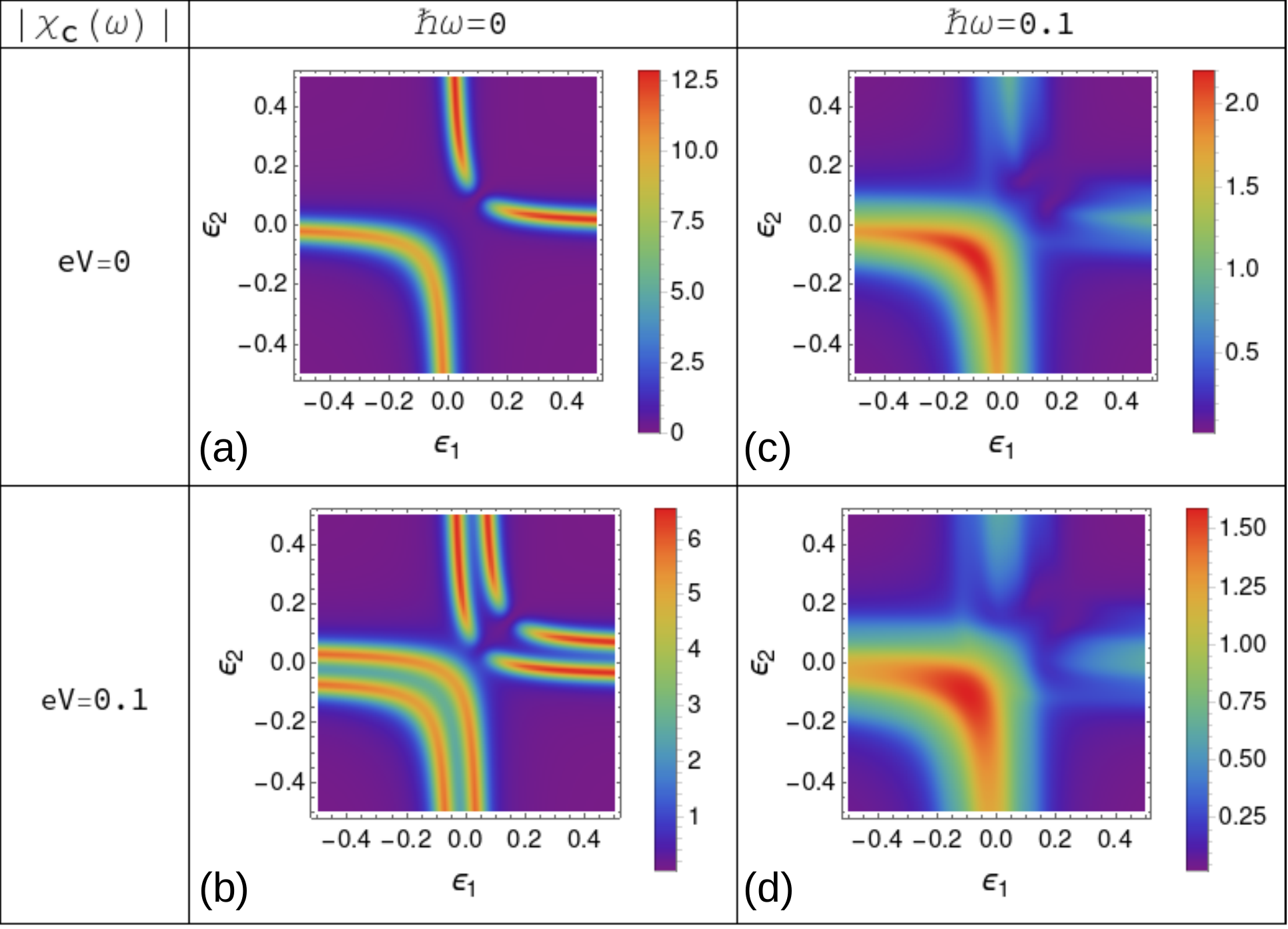}
\caption{Same as Fig.~\ref{figure1} for a parallel DQD.}\label{figure2}
\end{center}
\end{figure}

 Figure~\ref{figure2} shows the results obtained for a parallel DQD with symmetrical couplings: $\Gamma_{\alpha,ij}\equiv \Gamma$, $\forall \alpha, i, j$, with $\alpha_1=\alpha_2=1$ and $\mathcal{V}_{12}=\mathcal{V}_{21}$. At zero frequency, one observes three differences towards the situation in series: (i)~the intensity of  $|\mathcal{X}_c(0)|$ is reduced on the branches $\mathcal{B}_+^{L,R}$; (ii)~$|\mathcal{X}_c(0)|$ undergoes an extinction of its intensity at the intersections between the branches $\mathcal{B}_-^{L,R}$ with the diagonal $\mathcal{D}$; (iii)~at finite voltage, $|\mathcal{X}_c(0)|$ is of equal intensity along both halves of the branches $\mathcal{B}_-^{L,R}$ (respectively $\mathcal{B}_+^{L,R}$). These differences are understood as follows. According to Eq.~(\ref{def_lambda}), one~has 
\begin{eqnarray}
\lambda_\pm= \frac{1}{2}\left(\varepsilon_1+\varepsilon_2-2i\Gamma\pm\sqrt{(\varepsilon_1-\varepsilon_2)^2+4 (\mathcal{V}_{12}-i\Gamma)^2}\right).\nonumber\\
\end{eqnarray}

The imaginary parts of $\lambda_\pm$ depend on $\varepsilon_1$ and $\varepsilon_2$, contrary to what happens in the case in series where the imaginary parts of $\lambda_\pm$ were equal to $-\Gamma/2$ (see Eq.~(\ref{lambda_series})). Typically the imaginary part of~$\lambda_+$ for the parallel DQD is large along the branches~$\mathcal{B}_+^{L,R}$ explaining the fact that the intensity of $|\mathcal{X}_c(0)|$ on the latter branches are reduced (property (i)). Along the diagonal~$\mathcal{D}$ of equation $\varepsilon_1=\varepsilon_2=\varepsilon_0$, one has $\lambda_-=\varepsilon_0-\mathcal{V}_{12}$ whereas $\lambda_+=\varepsilon_0+\mathcal{V}_{12}-2i\Gamma$. The imaginary part of $\lambda_-$ is zero, meaning that the bonding state of the parallel DQD is disconnected from the leads, eliminating any dissipation effect through contacts to leads, and causing a significant reduction in the intensity of $|\mathcal{X}_c(0)|$ at the intersection of branches $\mathcal{B}_-^{L,R}$ and diagonal $\mathcal{D}$ (property (ii)). Finally, the property~(iii) can be understood as follows: at finite voltage $|\mathcal{X}_c(0)|$ is maximal on the branches~$\mathcal{B}_-^{L,R}$ with equal intensity along both halves of the branches since each dot are equally connected to $L$ and $R$ leads for parallel DQD. At finite frequency, Figs.~\ref{figure2}(c) and (d) show the broadening of the branches $\mathcal{B}_{\pm}^{L,R}$ and the widening of the gap in the branches $\mathcal{B}_-^{L,R}$. At $T=0$, $\varepsilon_1=\varepsilon_2$ and $\mathcal{V}_{21}=\mathcal{V}_{12}$, an exact expression for~$\mathcal{X}_c(\omega)$ is derived from Eq.~(\ref{chi_DQD_green}) in the parallel geometry\cite{SM}. It reads as $\mathcal{X}_c(\omega)=\sum_\pm\sum_{\alpha=L,R}\mathcal{X}_{\pm,\alpha}(\omega)$ where $\mathcal{X}_{-,\alpha}(\omega)=0$ and
\begin{eqnarray}\label{Xc_parallel}
\mathcal{X}_{+,\alpha}(\omega)=-\frac{2\Gamma}{h\omega(\hbar\omega+4i\Gamma)}
\ln\Big(1-\frac{h\omega(\hbar\omega+4i\Gamma)}{4\Gamma}A_+(\mu_\alpha)\Big).\nonumber\\
\end{eqnarray}

%%%%%%%%%%%%%%%%%%%%%%%%%%%%%%%%%%%%%%%%%%%%%%%%%%%%%%%%%%%%%%%%%%
%																 %
%																 %
%		RC-CIRCUIT       										 %
%																 %
%																 %
%%%%%%%%%%%%%%%%%%%%%%%%%%%%%%%%%%%%%%%%%%%%%%%%%%%%%%%%%%%%%%%%%%

\section{Equivalent quantum RC-circuit}

\begin{figure}[t]
\begin{center}
\includegraphics[width=4cm]{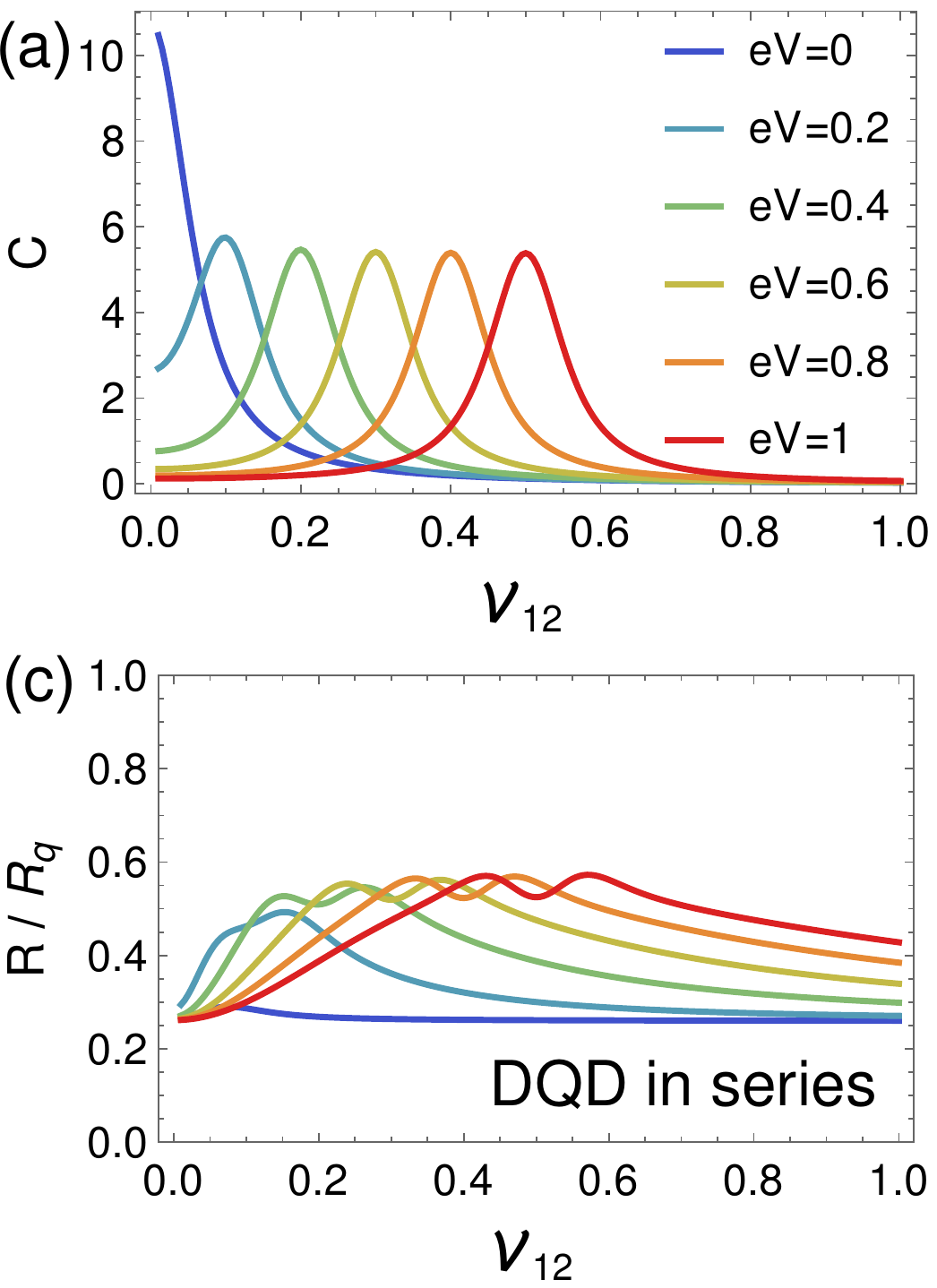}
\includegraphics[width=4cm]{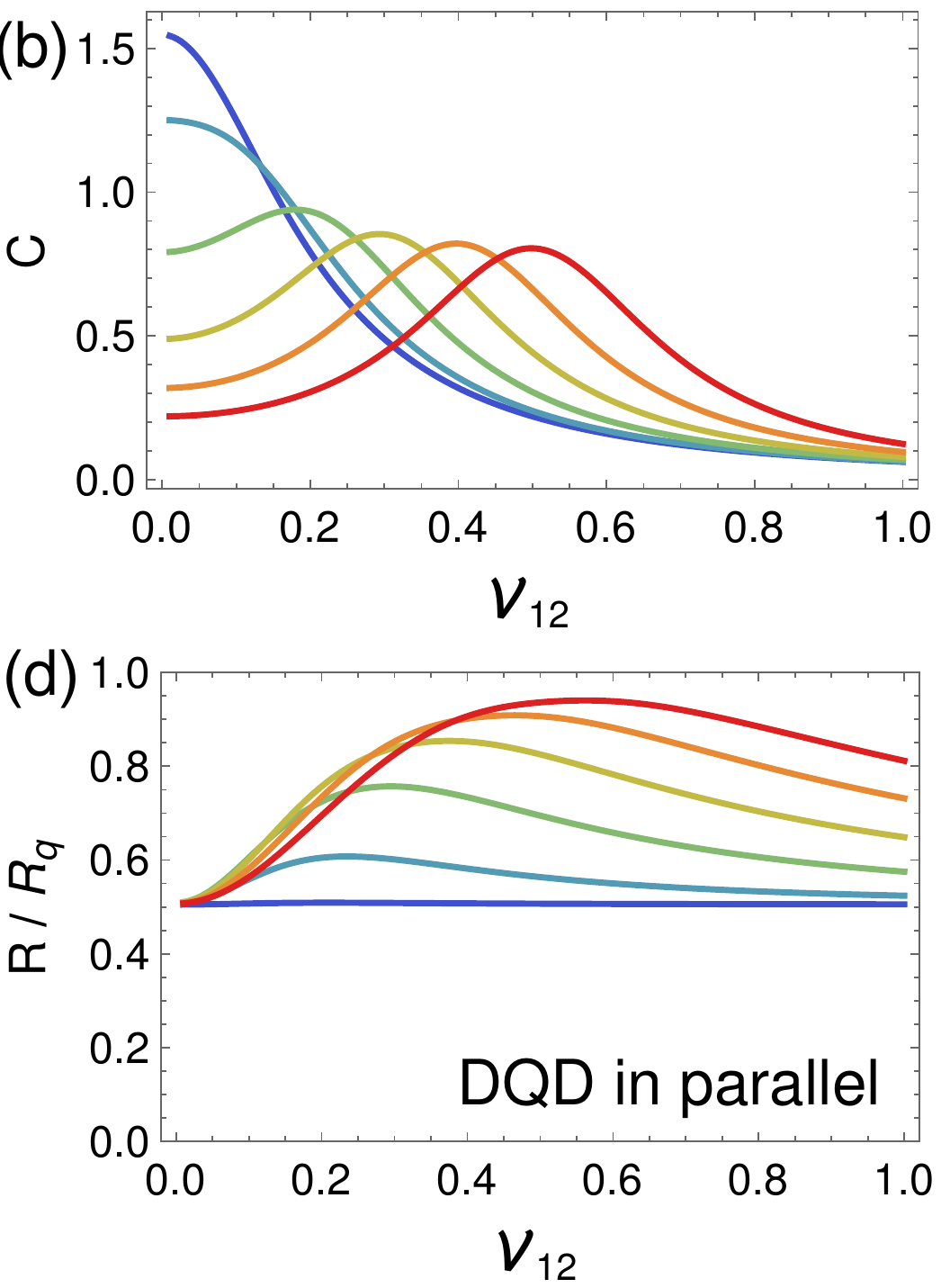}
\caption{Capacitance~$C$ and resistance~$R$ as a function of $\mathcal{V}_{12}$ for a DQD in (a), (c) series and (b), (d) parallel at various values of $V$ and $\varepsilon_1=\varepsilon_2=0$, $\Gamma=0.1$, $k_BT=0.01$.}\label{figureCR}
\end{center}
\end{figure}

\begin{figure}[t]
\begin{center}
\includegraphics[width=4cm]{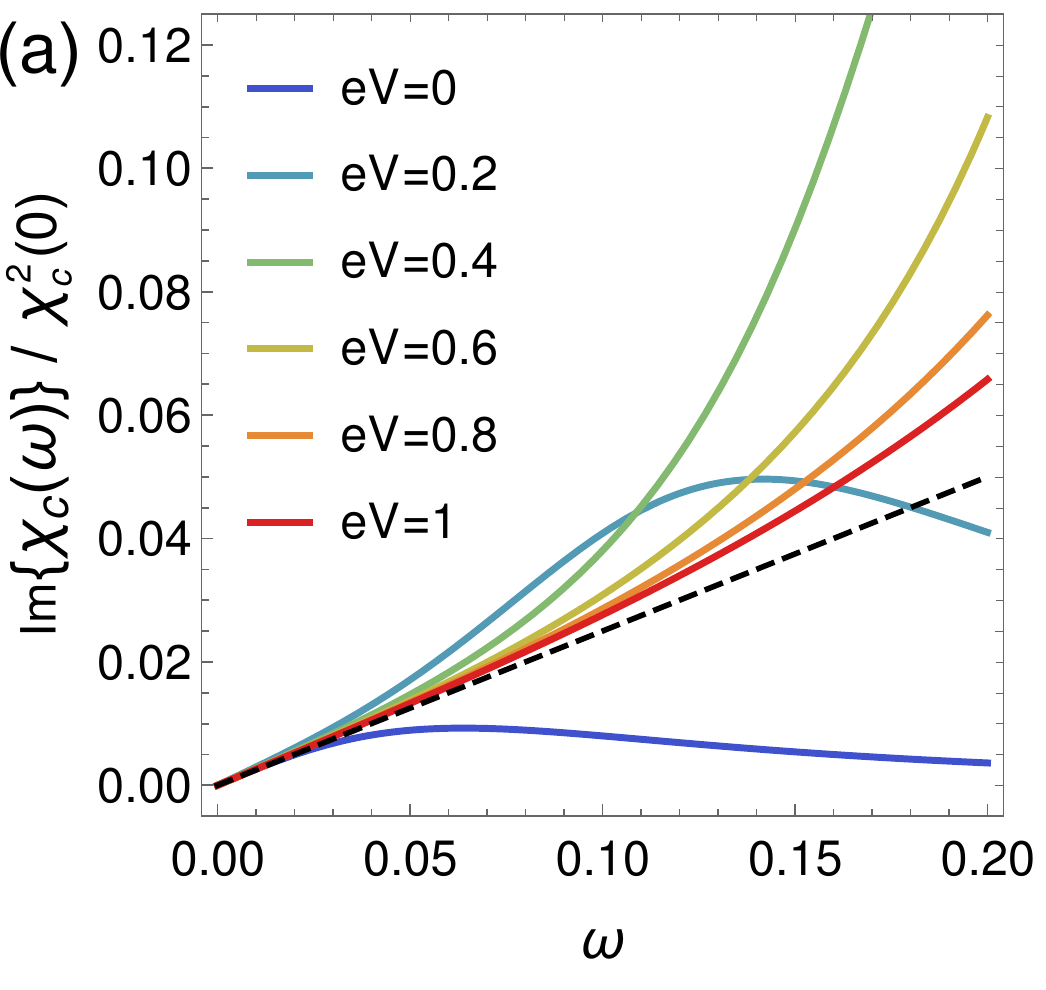}
\includegraphics[width=4cm]{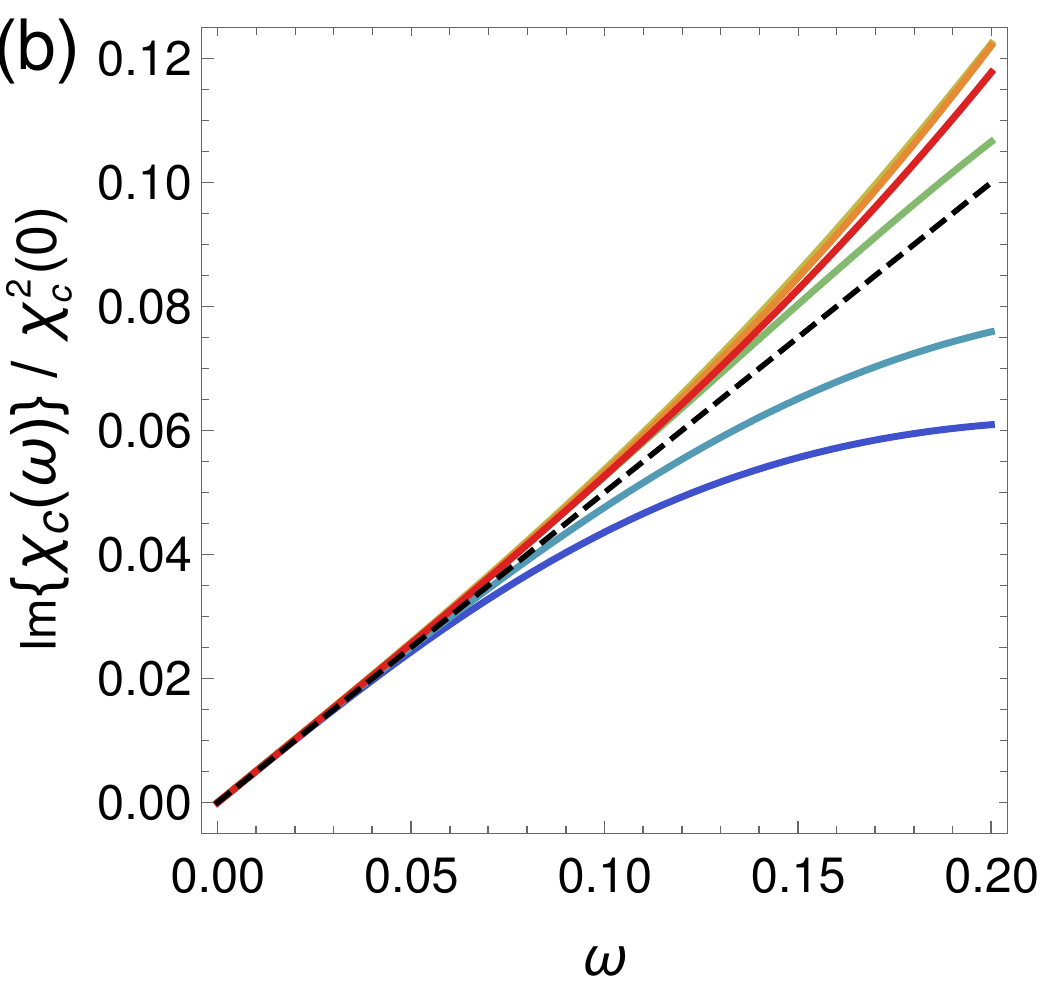}
\caption{$\text{Im}\{\mathcal{X}_c(\omega)\}/\mathcal{X}_c^2(0)$ as a function of $\omega$ for a DQD in (a)~series and (b) parallel at various values of $V$ and $\varepsilon_1=\varepsilon_2=0$, $\Gamma=0.1$, $\mathcal{V}_{12}=0.01$, $k_BT=0.01$. The equation of the dashed line is $\omega/4$ for the serial DQD and $\omega/2$ for the parallel~DQD.}\label{figureLF}
\end{center}
\end{figure}

We now focus on the characterization of the equivalent quantum RC-circuit to the DQD whose capacitance and charge relaxation resistance are respectively given by  $C=e^2\mathcal{X}_c(0)$ and $R=e^2\lim_{\omega\rightarrow 0}\text{Im}\{\mathcal{X}_c(\omega)\}/(\omega C^2)$\cite{Buttiker1993,Buttiker1993prl,Filippone2013}. To analyze them, we report in Fig.~\ref{figureCR} their values, extracted from $\mathcal{X}_c(\omega)$, versus the interdot coupling~$\mathcal{V}_{12}$. For a serial DQD, one observes that $C$ is maximal at $\mathcal{V}_{12}=eV/2$. Strictly speaking, $C$ diverges towards infinity at $V=T=0$ when $\mathcal{V}_{12}$ tends to 0 since the two dots are decoupled. For a parallel DQD, one has a reduction of amplitude for $C$ and the absence of a divergence when $\mathcal{V}_{12}$ tends to 0 at $V=T=0$ because there is always a way for the charges to travel from one lead to the other, even at $\mathcal{V}_{12}=0$. As far as the charge relaxation resistance~$R$ is concerned, the results reported in Figs.~\ref{figureCR}(c) and (d) show that its value at $\mathcal{V}_{12}=0$ equals $R_q/4$ in the case in series while it equals $R_q/2$ in the case in parallel, where $R_q=h/e^2$ is the quantum of resistance.  In both cases, $R$ versus $\mathcal{V}_{12}$ increases and then converges back to $R_q/4$, respectively $R_q/2$, at strong $\mathcal{V}_{12}$. To explain why the variation ranges of $R$ differ so markedly depending whether the DQD is connected in series or in parallel, let us start from Eqs.~(\ref{Xc_series}) and~(\ref{Xc_parallel}), which give $\mathcal{X}_c(\omega)$ at $T=0$ and $\varepsilon_1=\varepsilon_2$,  respectively for a serial and a parallel DQD symmetrically coupled to the leads. For a serial DQD, Eq.~(\ref{Xc_series}) leads to $C=(e^2/2)\sum_\pm\sum_{\alpha=L,R}A_\pm(\mu_\alpha)$ and
\begin{eqnarray}\label{Rq_series}
 R=\frac{\sum_\pm\sum_{\alpha=L,R}A^2_\pm(\mu_\alpha)}{(\sum_\pm\sum_{\alpha=L,R}A_\pm(\mu_\alpha))^2}R_q~.
\end{eqnarray}
By putting the latter expression in the form $(\sum_{i=1}^n x_i^2)(\sum_{i=1}^n y_i^2)R_q/(\sum_{i=1}^n x_iy_i)^2$ with $x_i=A_\pm(\mu_\alpha)$, $y_i=1$, and $n=4$, and using the Cauchy-Schwarz inequality, one deduces that $R\ge R_q/4$. Moreover, by knowing that $\sum_{i=1}^n z_i^4\le (\sum_{i=1}^n z_i^2)^2$ with $z_i=\sqrt{A_\pm(\mu_\alpha)}$, one concludes that  $R\le R_q$\cite{SM}. Thus the range of variation for $R$ is from $R_q/4$ to $R_q$, in line with what is observed in Fig.~\ref{figureCR}(c). For a parallel DQD, the expression for $\mathcal{X}_c(\omega)$ given by Eq.~(\ref{Xc_parallel}) leads to $C=(e^2/2)\sum_{\alpha=L,R}A_+(\mu_\alpha)$ and
\begin{eqnarray}\label{Rq_parallel}
 R=\frac{\sum_{\alpha=L,R}A^2_+(\mu_\alpha)}{(\sum_{\alpha=L,R}A_+(\mu_\alpha))^2}R_q~.
\end{eqnarray}
By employing similar arguments to those used for a serial DQD, and by noticing that $n=2$ in the case in parallel since the bonding state is disconnected from the leads at $\varepsilon_1=\varepsilon_2$, one shows that $R$ varies from  $R_q/2$ to $R_q$, in agreement with what is observed in Fig.~\ref{figureCR}(d). It means that when $\varepsilon_1=\varepsilon_2$, the number of quantum channels $n$ allowing the charge to travel are equal to four in the case in series whereas it equals two in the case in parallel. The results given by Eqs.~(\ref{Rq_series}) and~(\ref{Rq_parallel}) can be viewed as the extension to a DQD of the results previously established in the cases of a single quantum dot\cite{Pretre1995} or a quantum point contact\cite{Gabelli2006,Nigg2006,Mora2010,Gabelli2012,Filippone2020,Duprez2021}. As in such systems, $\mathcal{X}_c(\omega)$ in a DQD obeys a generalized Korringa-Shiba relation\cite{Garst2005,Filippone2012} according to which
$\lim_{\omega\rightarrow 0}\text{Im}\{\mathcal{X}_c(\omega)\}/\omega =\sum_{\pm}\sum_{\alpha=L,R}\mathcal{X}_{\pm,\alpha}^2(0)$,
a characteristic of a Fermi liquid. It is worthwhile to explore the variation of $\mathcal{X}_c(\omega)$ at higher frequencies and to see how its frequency dependence deviates from this relation. Figure~\ref{figureLF} reports the results obtained for $\text{Im}\{\mathcal{X}_c(\omega)\}/\mathcal{X}_c^2(0)$ as a function of $\omega$. For a serial DQD, respectively parallel DQD, one observes a linear variation with frequency according to $\omega/4$, respectively $\omega/2$, at low frequencies in agreement with the generalized Korringa-Shiba relation, confirming the difference of a factor two found for the values of~$R$ between the cases in series and in parallel, whereas it shows strong deviations at higher frequencies. 

%%%%%%%%%%%%%%%%%%%%%%%%%%%%%%%%%%%%%%%%%%%%%%%%%%%%%%%%%%%%%%%%%%
%																 %
%																 %
%		REFELECTION PHASE  											 %
%																 %
%																 %
%%%%%%%%%%%%%%%%%%%%%%%%%%%%%%%%%%%%%%%%%%%%%%%%%%%%%%%%%%%%%%%%%%

\section{Reflection phase}

\begin{figure}[b]
\begin{center}
\includegraphics[height=3.4cm]{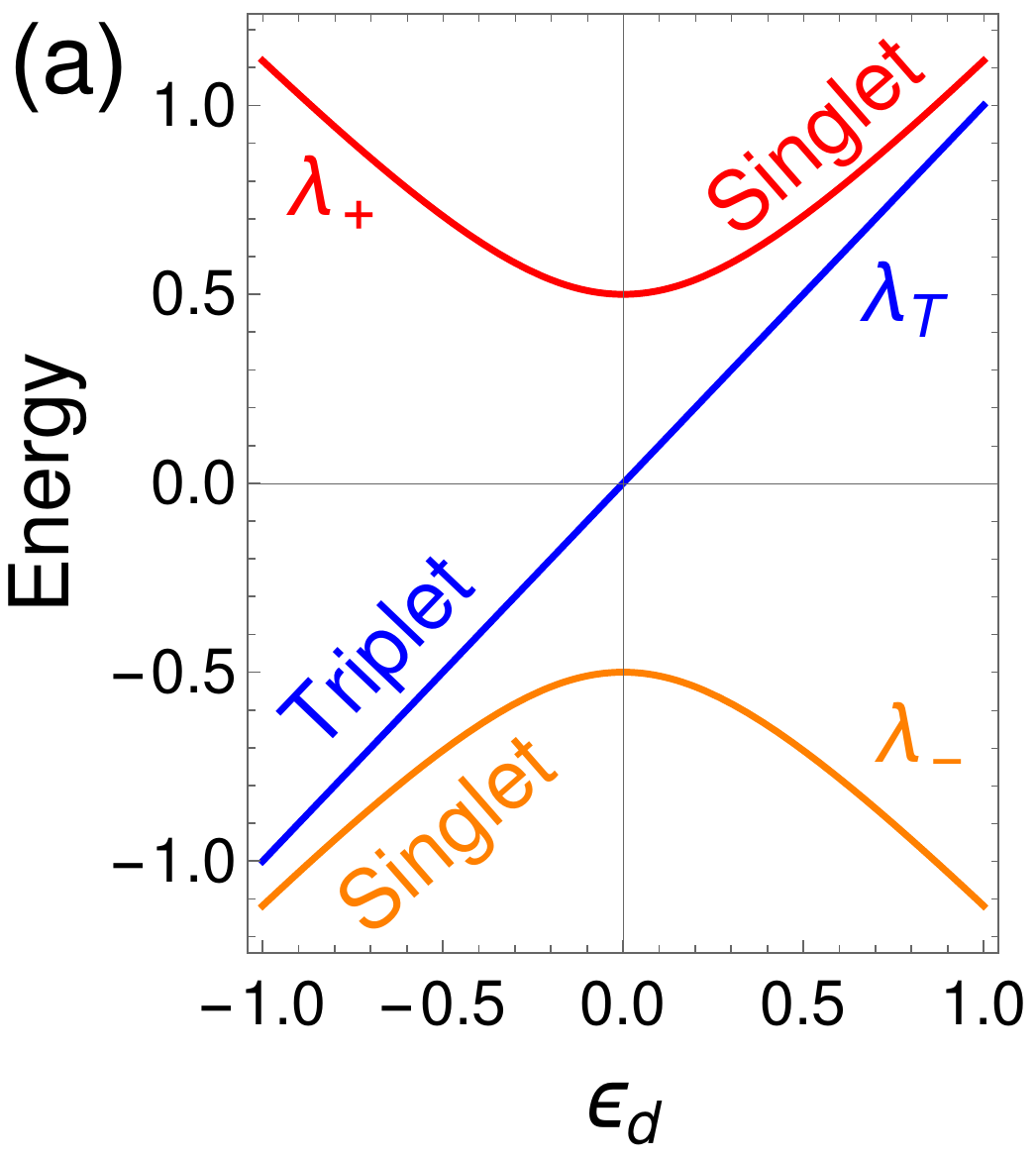}
\includegraphics[height=3.5cm]{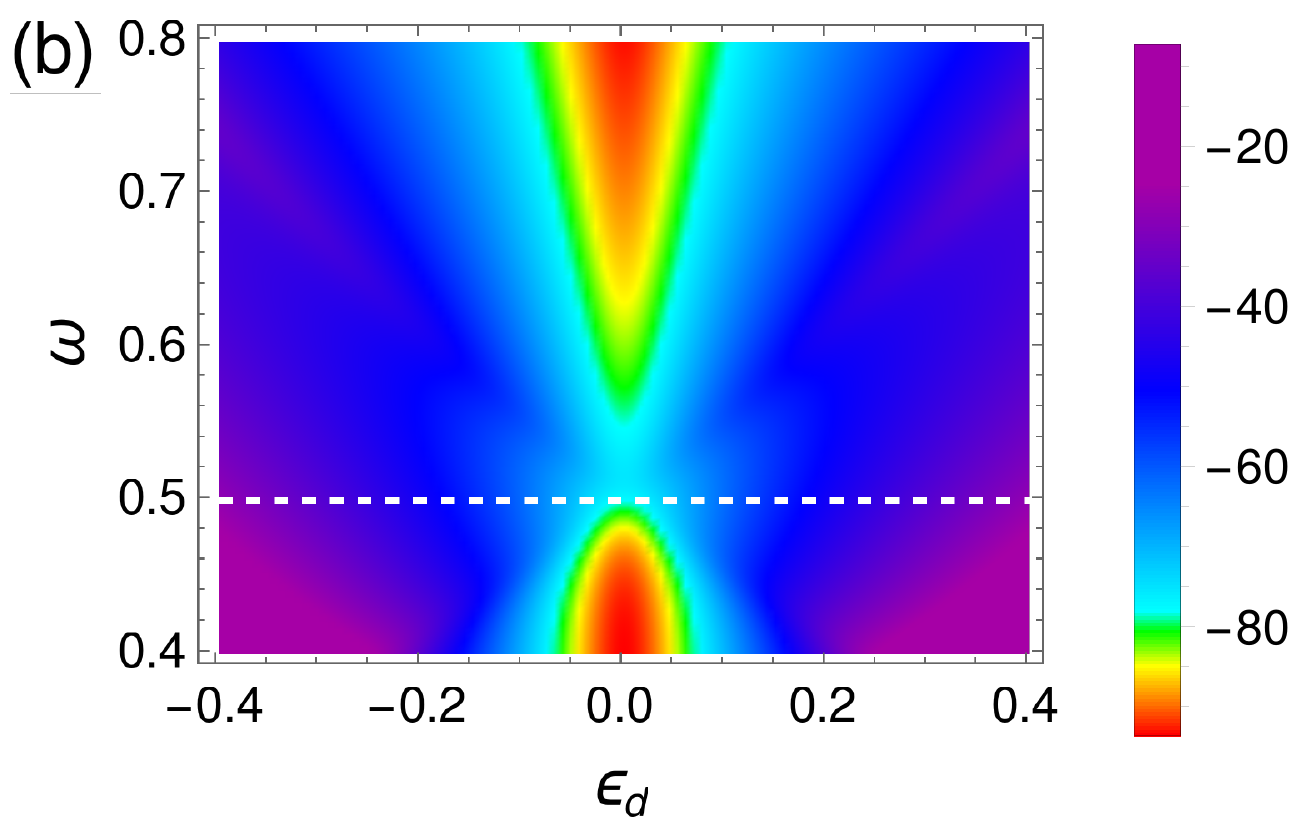}
\caption{(a) $\varepsilon_d$-dispersion of the singlet and triplet eigenvalues at $\varepsilon_1+\varepsilon_2=0$. (b) Color-scale plot of  $\phi(\omega)$ in mrad as a function of $\varepsilon_d$ and $\omega$ at $\varepsilon_1=T=V=0$, $\Gamma=0.04$, and $\mathcal{V}_{12}=0.5$ for a serial DQD. The dashed line corresponds to $\hbar\omega=\mathcal{V}_{12}$.}\label{figurephase}
\end{center}
\end{figure}

We end up by discussing the reflection phase of the system considered as a resonator embedded in an electromagnetic environment, which is defined as the phase shift between incoming and reflected microwaves, as measured in reflectometry experiments\cite{Schroer2012,Chorley2012,Colless2013,Viennot2014,Ezzouch2021}. A rapid calculation shows that this phase is related to the DCS via the relation $\phi(\omega)=\arctan(\text{Re}\{\mathcal{X}_c(\omega)\}/\text{Im}\{\mathcal{X}_c(\omega)\})$\cite{SM}. To compare our predictions to the measurements performed in spin qubit systems, we generalize the description of the spinless DQD in series made above by taking spin into account. This is simply done by considering triplet states in addition to the bonding and anti-bonding states which correspond to the singlet states in the case of a DQD with spin. The eigenenergy of the triplet state is given by  $\lambda_T=(\varepsilon_d-i\Gamma)/2$, where $\varepsilon_d=\varepsilon_2-\varepsilon_1$ is the detuning energy\cite{Petta2005}. Figure~\ref{figurephase}(a) shows the $\varepsilon_d$-dispersion of the eigenenergies $\lambda_\pm$ and $\lambda_T$. At $T=0$, we use a generalized expression for $\mathcal{X}_c(\omega)$ obtained from Eq.~(\ref{Xc_series}) by adding a term corresponding to the triplet state contribution according to $-3\Gamma/(2h\omega(\hbar\omega+i\Gamma))\sum_{\alpha=L,R}
 \ln(1-h\omega(\hbar\omega+i\Gamma)A_T(\mu_\alpha)/\Gamma)$, where $A_T(\varepsilon)=-\text{Im}\{\mathcal{G}^r_T(\varepsilon)\}/\pi$ with $\mathcal{G}^r_T(\varepsilon)=(\varepsilon-\lambda_T)^{-1}$. The results for $\phi(\omega)$ as a function of detuning energy and frequency is shown in Fig.~\ref{figurephase}(b). One observes a dip in phase inside two pockets spreading out symmetrically around the vertical axis $\varepsilon_d=0$, the two pockets being separated by a gap area located around $\hbar\omega=\mathcal{V}_{12}$, spotted by the dashed line. The existence of this gap is a direct consequence of the presence of bonding and anti-bonding states whereas the formation of the low-frequency pocket below the gap results from the presence of the triplet state. Our prediction for $\phi(\omega)$ is in good qualitative agreement with the experimental results obtained in spin qubits\cite{Ezzouch2021}.

%%%%%%%%%%%%%%%%%%%%%%%%%%%%%%%%%%%%%%%%%%%%%%%%%%%%%%%%%%%%%%%%%%
%																 %
%																 %
%		CONCLUSION  											 %
%																 %
%																 %
%%%%%%%%%%%%%%%%%%%%%%%%%%%%%%%%%%%%%%%%%%%%%%%%%%%%%%%%%%%%%%%%%%

\section{Conclusion} 

We have developed a model to calculate the DCS in a nonequilibrium DQD. We have established a general expression for this quantity, which, at $T=0$ is related to the spectral function contributions from the bonding and anti-bonding states, leading to the prediction of a splitting of the two branches of maxima of the DCS as a function of the energy levels of the dots resulting from the application of a finite bias voltage driving the DQD out-of-equilibrium. It would be interesting to check this prediction experimentally. In the low frequency regime, we have analyzed the results in terms of the capacitance and the resistance of the equivalent RC-circuit in both serial and parallel geometries. In the high frequency regime, by extending our results to take into account an additional triplet mode in order to describe spin qubits, we have deduced the evolution for~$\mathcal{X}_c(\omega)$ as a function of~$\omega$ and~$\varepsilon_d$ and found a qualitative agreement with experimental results recently obtained. The approach presented here can be used in many other contexts: quantum dots with multiple energy levels, submitted to the application of a magnetic field, in the presence of exchange or Coulomb interactions\cite{Lopez2002,Schroer2006,Kashcheyevs2007,Juergens2013} among others.

\acknowledgments
We dedicate this article to Marc Sanquer$\dagger$. We pay tribute to his memory for his role in the development of quantum nanoelectronics. Marc had played a major role in our interest in this topics and never ceased to provide us with encouragement, sharing with us his expertise and enthusiasm for this field of research. We would like to thank Romain Maurand for valuable discussions and help and Vyacheslavs Kashcheyevs for stimulating discussion.

%%%%%%%%%%%%%%%%%%%%%%%%%%%%%%%%%%%%%%%%%%%%%%%%%%%%%%%%%%%%%%%%%%
%																 %
%																 %
%		REFERENCES											     %
%																 %
%																 %
%%%%%%%%%%%%%%%%%%%%%%%%%%%%%%%%%%%%%%%%%%%%%%%%%%%%%%%%%%%%%%%%%%

\bibliographystyle{apsrev4-1}
%\bibnote[]{} pour inserer une note dans le texte.

%%%%%%%%%%%%%%%%%%%%%%%%%%%%%%%%%%%%%%%%%%%%%%%%%%%%%%%%%%%%%%%%%%
%																 %
%																 %
%		SM       											     %
%																 %
%																 %
%%%%%%%%%%%%%%%%%%%%%%%%%%%%%%%%%%%%%%%%%%%%%%%%%%%%%%%%%%%%%%%%%%

\widetext

\pagebreak

\begin{center}
\textbf{Dynamical charge susceptibility in nonequilibrium double quantum dots\\ -- Supplemental Material --}\\~\\
A. Cr\'epieux$^1$ and M. Lavagna$^{2,3}$\\
{\small\it $^1$Aix Marseille Univ, Universit\'e de Toulon, CNRS, CPT, Marseille, France\\
$^2$ Univ. Grenoble Alpes, CEA, IRIG, PHELIQS, F-38000 Grenoble, France\\
$^3$ Centre National de la Recherche Scientifique, CNRS, 38042 Grenoble, France}\\
\end{center}

%%%%%%%%%% Merge with supplemental materials %%%%%%%%%%
%%%%%%%%%% Prefix a "S" to all equations, figures, tables and reset the counter %%%%%%%%%%
\setcounter{equation}{0}
\setcounter{figure}{0}
\setcounter{table}{1}
\setcounter{page}{1}
\makeatletter
\renewcommand{\theequation}{S\arabic{equation}}
\renewcommand{\thefigure}{S\arabic{figure}}
%%%%%%%%%% Prefix a "S" to all equations, figures, tables and reset the counter %%%%%%%%%%

In this Supplemental Material, we give the details of the calculation concerning (i)~the integration over Grassmann variables, (ii)~the dynamical charge susceptibility $\mathcal{X}_c(\omega)$ for both a DQD in series and in parallel, (iii)~the analytical expression for $\mathcal{X}_c(\omega)$ in the limit of zero temperature ($T=0$), (iv)~the capacitance~$C$ and the charge relaxation resistance~$R$ of the equivalent RC-circuit for aligned dot levels $\varepsilon_1=\varepsilon_2$, (v)~the determination of the phase response $\phi(\omega)$ of the corresponding resonator, and (vi)~the expression for $\mathcal{X}_c(\omega)$ in a DQD in series in the presence of additional triplet states.

\tableofcontents

%%%%%%%%%%%%%%%%%%%%%%%%%%%%%%%%%%%%%%%%%%%%%%%%%%%%%%%%%%%%%%%%%%
%																 %
%																 %
%		INTEGRATION ON THE LEADS           						 %
%																 %
%																 %
%%%%%%%%%%%%%%%%%%%%%%%%%%%%%%%%%%%%%%%%%%%%%%%%%%%%%%%%%%%%%%%%%%

\section{Function integral approach}

%{\color{red} Attention, il y a une ambiguite entre variables de Grassmann et operateurs dans la section suivante...}

Within the functional integral approach, the partition function can be written as the integral over the Grassmann variables $d_i^\dg$, $d_i$, $c_{\alpha k}^\dg$, $c_{\alpha k}$, associated with the creation and annihilation operators $\widehat d_i^\dg$, $\widehat d_i$ for the dots, and $\widehat c_{\alpha k}^\dg$, $\widehat c_{\alpha k}$ for the leads. Indeed, one has
\begin{eqnarray}
 \mathcal{Z}=\int \prod_{i=1,2}\mathcal{D}d_i\prod_{\substack{\a =L,R\\k \in \a}}\mathcal{D}c_{\a k}\exp\left(-\int_0^{\beta}d\tau\mathcal{L}(\tau)\right),
\end{eqnarray}
where $\beta=1/k_BT$, $\mathcal{D}d_{i}$ and $\mathcal{D}c_{\alpha k}$ are the differential elements defined as $\mathcal{D}d_{i}=\text{d}d_{i}^\dg \text{d}d_{i}$ and $\mathcal{D}c_{\alpha k}= \text{d}c_{\alpha k}^\dg  \text{d}c_{\alpha k}$. $\mathcal{L}(\tau)$~is the Lagrangian of the system defined in three parts  $\mathcal{L}(\tau)=\mathcal{L}_\text{dots}(\tau)+\mathcal{L}_\text{leads}(\tau)+\mathcal{L}_\text{hop}(\tau)$, where
$\mathcal{L}_\text{dots}(\tau)=\sum_{i=1,2}d_{i}^\dg\partial_\tau d_{i}-\widehat {\m H}_\text{dots}$,
$ \mathcal{L}_\text{leads}(\tau)=\sum_{\substack{\a =L,R ; k \in \a}}c_{\alpha k}^\dg\partial_\tau c_{\alpha k}-\widehat {\m H}_\text{leads}$ and
$\mathcal{L}_\text{hop}(\tau)=-\widehat {\m H}_\text{hop}$. One remarks that the sum of $\mathcal{L}_\text{leads}(\tau)$ and $\mathcal{L}_\text{hop}(\tau)$ can be written equivalently in the form of a perfect square from which a supplementary term is subtracted
\begin{eqnarray}
\mathcal{L}_\text{leads}(\tau)+ \mathcal{L}_\text{hop}(\tau)=
\sum_{\substack{\a =L,R\\k \in \a}}\widetilde c_{\alpha k}^\dg(\partial_\tau+\varepsilon_{\alpha k})\widetilde c_{\alpha k}
-\sum_{\substack{\a =L,R ; k \in \a}}\sum_{i,j}V_{i,\alpha k}^* (\partial_\tau+\varepsilon_{\alpha k})^{-1}V_{j,\alpha k}d_i^\dg d_j,
\end{eqnarray}
with $\widetilde c_{\alpha k}=c_{\alpha k}-\sum_{j=1,2}V_{j,\alpha k}(\partial_\tau+\varepsilon_{\alpha k})^{-1}d_j$. By integrating over the Grassmann variables $\widetilde c_{\alpha k}$ and $\widetilde c_{\alpha k}^\dg$, one gets 
$\mathcal{Z}=\int \mathcal{D}d_1\mathcal{D}d_2\exp(-\int_0^{\beta}d\tau\mathcal{L}_\text{eff}(\tau))$,
within a constant multiplicative factor. $\mathcal{L}_\text{eff}(\tau)$ is an effective Lagrangian defined as
\begin{eqnarray}
 &&\mathcal{L}_\text{eff}(\tau)=
\left(
\begin{array}{c}
d_1^\dg\;\;d_2^\dg
\end{array}
\right)\nonumber\\
&&\times\left(
\begin{array}{cc}
\partial_\tau+\varepsilon_1+{\mathbb{\Sigma}}_{11}(\tau)& \widetilde{\mathbb{\Sigma}}_{12}(\tau)\\
&\\
\widetilde{\mathbb{\Sigma}}_{21}(\tau)&\partial_\tau+\varepsilon_2+{\mathbb{\Sigma}}_{22}(\tau)
\end{array}
\right)
\left(
\begin{array}{c}
d_1\\
\\
d_2
\end{array}
\right),
\end{eqnarray}
where ${\mathbb{\Sigma}}_{ij}(\tau)=\sum_{\a =L,R}\sum_{k \in \a}V_{i,\alpha k}^*g_{\alpha k}(\tau) V_{j,\alpha k}$, with $g_{\a k}(\tau)=-(\partial_\tau+\varepsilon_{\alpha k})^{-1}$ and $\widetilde {\mathbb{\Sigma}}_{i\overline{i}}(\varepsilon)={\mathbb{\Sigma}}_{i\overline{i}}(\varepsilon)+{\m V}_{i\overline{i}}^*$. The retarded Green function in the dots can be obtained from
$G^r_{ij}(\tau)=\int \mathcal{D}d_1\mathcal{D}d_2 d_i d_j^\dg \exp(-\int_0^{\beta}d\tau\mathcal{L}_\text{eff}(\tau))$\cite{Negele1988sm}.
By performing the integration over the Grassmann variables $d_i$ and $d_i^\dg$, and taking the Fourier transform, one gets
\begin{eqnarray}\label{green_function}
\doubleunderline G^r(\varepsilon)=\left(
\begin{array}{cc}
\varepsilon-\varepsilon_1-{\mathbb{\Sigma}}^r_{11}(\varepsilon)& -\widetilde{\mathbb{\Sigma}}^r_{12}(\varepsilon)\\
&\\
-\widetilde{\mathbb{\Sigma}}^r_{21}(\varepsilon)&\varepsilon-\varepsilon_2-{\mathbb{\Sigma}}^r_{22}(\varepsilon)
\end{array}
\right)^{-1}.
\end{eqnarray}
%After inverting the matrix, one gets a result for $\doubleunderline G^r(\varepsilon)$ which coincides with the one obtained by using the equation of motion approach\cite{Lavagna2020}.

From Eq.~\ref{green_function}, one extracts an effective Hamiltonian $\mathcal{H}_\text{eff}$ defined through the relation $ \doubleunderline G^r(\varepsilon)=(\vep-\mathcal{H}_\text{eff})^{-1}$. It is given by
\begin{eqnarray}
\mathcal{H}_\text{eff}=\left(
\begin{array}{ccc}
\varepsilon_1-i\cfrac{\Gamma_{11}}{2}& & \mathcal{V}_{12}^{\,*}-i\cfrac{\Gamma_{12}}{2}\\
& &\\
\mathcal{V}_{21}^{\,*}-i\cfrac{\Gamma_{21}}{2}& &\varepsilon_2-i\cfrac{\Gamma_{22}}{2}
\end{array}
\right),
\end{eqnarray}
since $\mathbb{\Sigma}^r_{ij}(\varepsilon)=-i\Gamma_{ij}/2$ and $\Gamma_{ij}=\sum_{\alpha=L,R}\Gamma_{\alpha,ij}$, where $\Gamma_{\alpha,ij}=2\pi V^*_{i,\alpha k}V_{j,\alpha k}\rho_\alpha$ does not depend on $k$  in the flat-wide-band approximation for electrons in the leads. The fact that $\mathcal{H}_\text{eff}$ is non-hermitian is the consequence of the integration over the lead Grassmann variables that has been performed. The coupling to the leads is thus explicitly taken into account in the Hamiltonian. It induces a dissipation\cite{Wong1967sm} since the eigenvalues of $\mathcal{H}_\text{eff}$ acquire an imaginary part
\begin{eqnarray}\label{def_lambda_general}
\lambda_\pm= \frac{1}{2}\left(\varepsilon_1-i\frac{\Gamma_{11}}{2}+\varepsilon_2-i\frac{\Gamma_{22}}{2}\pm \Delta\right),
\end{eqnarray}
with
$\Delta^2=(\varepsilon_1-i\Gamma_{11}/2-\varepsilon_2+i\Gamma_{22}/2)^2+4( \mathcal{V}_{12}^{\,*}-i\Gamma_{12}/2)(\mathcal{V}_{21}^{\,*}-i\Gamma_{21}/2)$. The eigenvalues $\lambda_-$ and $\lambda_+$ are respectively the energies of the bonding and anti-bonding states of the DQD system. The non-hermitian matrix $\doubleunderline U$, of elements $U_{ij}$, associated with this diagonalization is
\begin{eqnarray}\label{def_U}
\doubleunderline U=\left(
\begin{array}{cc}
\cfrac{\mathcal{V}_{12}^{\,*}-i\Gamma_{12}/2}{D_1} & \cfrac{\delta}{2D_2}\\
&\\
-\cfrac{\delta}{2D_1}&\cfrac{\mathcal{V}_{21}^{\,*}-i\Gamma_{21}/2}{D_2}
\end{array}
\right),
\end{eqnarray}
with 
\begin{eqnarray}
 \delta&=&\varepsilon_1-i\frac{\Gamma_{11}}{2}-\varepsilon_2+i\frac{\Gamma_{22}}{2}-\Delta,\\
 D_i^2&=&\left|\mathcal{V}_{i\overline{i}}^{\,*}-i\frac{\Gamma_{i\overline{i}}}{2}\right|^2+\frac{|\,\delta\,|^2}{4}.
\end{eqnarray}

%%%%%%%%%%%%%%%%%%%%%%%%%%%%%%%%%%%%%%%%%%%%%%%%%%%%%%%%%%%%%%%%%%
%																 %
%																 %
%		X(w) for a DQD                    						 %
%																 %
%																 %
%%%%%%%%%%%%%%%%%%%%%%%%%%%%%%%%%%%%%%%%%%%%%%%%%%%%%%%%%%%%%%%%%%

\section{General expression for the dynamical charge susceptibility}

\subsection{Calculation of $\mathcal{X}_c(\omega)$}

The charge susceptibility is given by $\mathcal{X}_c(\omega)=\sum_{i,j=1,2}\alpha_i\alpha_j\mathcal{X}_{ij}(\omega)$, where $\alpha_i$ are the level-arm coefficients measuring the asymmetry of the capacitive couplings\cite{Lavagna2020sm}, and $\mathcal{X}_{ij}(\omega)$ is the Fourier transform of
\begin{eqnarray}
 \mathcal{X}_{ij}(t,t')=i\Theta(t-t')\langle \{\Delta  \widehat N_j(t'),\Delta \widehat N_i(t)\}\rangle~,
\end{eqnarray}
where $\Delta \widehat N_i(t)= \widehat N_i(t)-\langle  \widehat N_i \rangle$ with $ \widehat N_i(t)= \widehat d^{\;\dag}_{i}(t) \widehat d_{i}(t)$, and $\{,\}$ denotes the anti-commutator. To calculate this quantity, one has first to calculate the time-ordered correlator defined as
\begin{eqnarray}
 \mathcal{X}^t_{ij}(\tau,\tau')&=&i\langle T_c[\Delta \widehat N_j(\tau')\Delta \widehat N_i(\tau)]\rangle\nonumber\\
&=&i\langle T_c[ \widehat N_j(\tau') \widehat N_i(\tau)]\rangle-i\langle T_c[ \widehat N_j]\rangle\langle T_c[ \widehat N_i]\rangle\nonumber\\
&=&i\langle T_c[ \widehat d^{\;\dag}_{j}(\tau') \widehat d_{j}(\tau') \widehat d^{\;\dag}_{i}(\tau) \widehat d_{i}(\tau)]\rangle
-i\langle T_c[ \widehat d^{\;\dag}_{j} \widehat d_{j}]\rangle\langle T_c[ \widehat d^{\;\dag}_{i} \widehat d_{i}]\rangle~,
\end{eqnarray}
where $T_c$ is the time-ordered operator. By assuming that there is no Coulomb interaction, one can use the Wick theorem to calculate this correlator, thus
\begin{eqnarray}
 \mathcal{X}^t_{ij}(\tau,\tau')&=&i\langle T_c[ \widehat d^{\;\dag}_{j}(\tau') \widehat d_{i}(\tau)]\rangle\langle T_c[ \widehat d_{j}(\tau') \widehat d^{\;\dag}_{i}(\tau)]\rangle~.
\end{eqnarray}
By introducing the time-ordered Green functions defined as $G^t_{ij}(\tau,\tau')=-i\langle T_c[ \widehat d^{\;\dag}_{j}(\tau') \widehat d_{i}(\tau)]\rangle$, one gets:
\begin{eqnarray}
 \mathcal{X}^t_{ij}(\tau,\tau')&=&iG^t_{ij}(\tau,\tau')G^t_{ji}(\tau',\tau)~.
\end{eqnarray}
In the basis of the eigenstates associated with the operators $\widehat d_+$ and $\widehat d_-$ in which $\mathcal{H}_\text{eff}$ is diagonal, one has
\begin{eqnarray}
 \widehat d_1(t)=U_{11}\widehat d_+(t)+U_{12}\widehat d_-(t)~,\\
  \widehat d_2(t)=U_{21}\widehat d_+(t)+U_{22}\widehat d_-(t)~,\\
 \widehat d^{\;\dag}_1(t)=U_{11}^*\widehat d^{\;\dag}_+(t)+U_{12}^*\widehat d^{\;\dag}_-(t)~,\\
  \widehat d^{\;\dag}_2(t)=U_{21}^*\widehat d^{\;\dag}_+(t)+U_{22}^*\widehat d^{\;\dag}_-(t)~,
\end{eqnarray}
where $U_{ij}$ are the elements of the transition matrix $ \doubleunderline{U}$. Indeed, one has $\widehat d_i=\sum_{s=\pm}\langle i|s\rangle \widehat d_s$, with $\langle i|s\rangle=U_{is}$. It leads~to
\begin{eqnarray}
 G^t_{ij}(\tau,\tau')&=&U_{i1}U^*_{j1}\mathcal{G}^t_{+}(\tau,\tau')+U_{i2}U^*_{j2}\mathcal{G}^t_{-}(\tau,\tau')~,\\
 G^t_{ji}(\tau',\tau)&=&U_{j1}U^*_{i1}\mathcal{G}^t_{+}(\tau',\tau)+U_{j2}U^*_{i2}\mathcal{G}^t_{-}(\tau',\tau)~,
\end{eqnarray}
where $\mathcal{G}^t_{\pm}(\tau,\tau')=-i\langle T_c[ \widehat d^{\;\dag}_{\pm}(\tau') \widehat d_{\pm}(\tau)]\rangle$. Finally, one has
\begin{eqnarray}
 \mathcal{X}^t_{ij}(\tau,\tau')&=&i|U_{i1}U_{j1}|^2\mathcal{G}^t_{+}(\tau,\tau')\mathcal{G}^t_{+}(\tau',\tau)
 +iU_{i1}U^*_{i2}U^*_{j1}U_{j2}\mathcal{G}^t_{+}(\tau,\tau')\mathcal{G}^t_{-}(\tau',\tau)\nonumber\\
 &&+iU^*_{i1}U_{i2}U_{j1}U^*_{j2}\mathcal{G}^t_{-}(\tau,\tau')\mathcal{G}^t_{+}(\tau',\tau)
 +i|U_{i2}U_{j2}|^2\mathcal{G}^t_{-}(\tau,\tau')\mathcal{G}^t_{-}(\tau',\tau)~,
\end{eqnarray}
which leads to 
\begin{eqnarray}
\mathcal{X}^t_c(\tau,\tau')=\sum_{s_1,s_2=\pm}  \mathcal{C}_{s_1s_2} \mathcal{X}^t_{s_1s_2}(\tau,\tau')~,
\end{eqnarray}
where $ \mathcal{X}^t_{s_1s_2}(\tau,\tau')= i\mathcal{G}^t_{s_1}(\tau,\tau')\mathcal{G}^t_{s_2}(\tau',\tau)$
and $\mathcal{C}_{s_1s_2}$ are coherence factors defined as
\begin{eqnarray}
 \mathcal{C}_{++}&=&\sum_{i,j=1,2}\alpha_i\alpha_j|U_{i1}U_{j1}|^2~,\\
 \mathcal{C}_{+-}&=&\sum_{i,j=1,2}\alpha_i\alpha_jU_{i1}U^*_{i2}U^*_{j1}U_{j2}~,\\
 \mathcal{C}_{-+}&=&\sum_{i,j=1,2}\alpha_i\alpha_jU^*_{i1}U_{i2}U_{j1}U^*_{j2}~,\\
 \mathcal{C}_{--}&=&\sum_{i,j=1,2}\alpha_i\alpha_j|U_{i2}U_{j2}|^2~.
\end{eqnarray}
One underlines that $\mathcal{C}_{++}$ and $\mathcal{C}_{--}$ are both reals, and that $\mathcal{C}_{+-}^*=\mathcal{C}_{-+}$. 
By performing an analytical continuation\cite{Haug2010sm}, one obtains the charge susceptibility $\mathcal{X}_c(\tau,\tau')=\sum_{s_1,s_2=\pm}  \mathcal{C}_{s_1s_2} \mathcal{X}_{s_1s_2}(\tau,\tau')$ with
\begin{eqnarray}
 \mathcal{X}_{s_1s_2}(\tau,\tau')= i\Big[\mathcal{G}^<_{s_1s_2}(\tau,\tau')\mathcal{G}^a_{s_2}(\tau',\tau)
 +\mathcal{G}^r_{s_1}(\tau,\tau')\mathcal{G}^<_{s_1s_2}(\tau',\tau)\Big]~.
\end{eqnarray}
By taking the Fourier transform, one gets  $\mathcal{X}_c(\omega)=\sum_{s_1,s_2=\pm}  \mathcal{C}_{s_1s_2} \mathcal{X}_{s_1s_2}(\omega)$  with
\begin{eqnarray}\label{chi_DQD_green}
\boxed{\mathcal{X}_{s_1s_2}(\omega)=i\int_{-\infty}^\infty \frac{d\varepsilon}{2\pi} \Big[\mathcal{G}^<_{s_1s_2}(\varepsilon)\mathcal{G}^a_{s_2}(\varepsilon-\hbar\omega)+\mathcal{G}^r_{s_1}(\varepsilon+\hbar\omega)\mathcal{G}^<_{s_1s_2}(\varepsilon)\Big]}
\end{eqnarray}
The retarded and advanced Green functions $ \doubleunderline{\mathcal{G}}^{r,a}(\varepsilon)$ are diagonal matrices of elements $\mathcal{G}^r_{s}(\varepsilon)=(\varepsilon-\lambda_s)^{-1}$ and $\mathcal{G}^a_{s}(\varepsilon)=(\varepsilon-\lambda_s^*)^{-1}$ in the eigenstate basis associated with the operators $\widehat d_+$ and $\widehat d_-$, where $\lambda_s$ is given by Eq.~(\ref{def_lambda_general}). In the steady state, the nonequilibrium Green function is given by
\begin{eqnarray}
 \doubleunderline{\mathcal{G}}^<(\varepsilon)=\doubleunderline{\mathcal{G}}^r(\varepsilon)\doubleunderline U^{-1}\doubleunderline{\mathbb{\Sigma}}^<(\varepsilon)\,\doubleunderline U^{\,\text{ }}\doubleunderline{\mathcal{G}}^a(\varepsilon)~,
\end{eqnarray}
where $\doubleunderline{\mathbb{\Sigma}}^<(\varepsilon)=i\sum_{\alpha=L,R}f_\alpha(\varepsilon)\doubleunderline\Gamma_{\,\alpha}$ is the self-energy in the initial basis associated with the operators $d_1$ and $d_2$, $f_{\alpha}(\varepsilon)=[1+\exp((\varepsilon-\mu_{\alpha})/k_BT)]^{-1}$ is the Fermi-Dirac distribution functions of the left (L) and right (R) leads, $\mu_{L,R}$, the chemical potential, and $T$, the temperature. $\doubleunderline{\mathcal{G}}^<(\varepsilon)$ is usually a non-diagonal matrix in the eigenstate basis.

\subsection{Coherence factors $\mathcal{C}_{s_1s_2}$}

In case of symmetrical capacitive couplings, i.e. $\alpha_1=\alpha_2=1$, one has
\begin{eqnarray}
 \mathcal{C}_{++}&=&(|U_{11}|^2+|U_{12}|^2)^2=1~,\\
 \mathcal{C}_{+-}&=&\sum_{i,j=1,2}U_{i1}U^*_{i2}U^*_{j1}U_{j2}=\mathcal{C}_{-+}~,\\
  \mathcal{C}_{--}&=&(|U_{21}|^2+|U_{22}|^2)^2=1~.
\end{eqnarray}

In each of the possible geometries of the DQD, it leads to:
\begin{itemize}
 \item In the case of a DQD in series with symmetrical real couplings, one has $\Gamma_{L,11}=\Gamma_{R,22}\equiv \Gamma$, $\Gamma_{L,22}=\Gamma_{R,11}=0$, $\Gamma_{\alpha,12}=\Gamma_{\alpha,21}=0$, $\forall \alpha$, and $\mathcal{V}_{12}=\mathcal{V}_{21}$:
\begin{eqnarray}
 \mathcal{C}_{+-}=\mathcal{C}_{-+}=0~,
\end{eqnarray}
so that only $\mathcal{X}_{++}(\omega)$ and $\mathcal{X}_{--}(\omega)$ contribute to the charge susceptibility according to $\mathcal{X}_c(\omega)=\mathcal{X}_{++}(\omega)+\mathcal{X}_{--}(\omega)$.
\item In the case of a DQD in parallel with symmetrical real couplings, one has $\Gamma_{\alpha,ij}\equiv \Gamma$, $\forall \alpha, i, j$ and~$\mathcal{V}_{12}=\mathcal{V}_{21}$:
\begin{eqnarray}\label{Cpm}
 \mathcal{C}_{+-}=\mathcal{C}_{-+}=\frac{|(\mathcal{V}_{12}-i\Gamma)\delta|^2-\text{Re}\{(\mathcal{V}_{12}+i\Gamma)^2\delta^2\}}{2D^4}~,
\end{eqnarray}
where $D^2=|\mathcal{V}_{12}-i\Gamma|^2+|\delta|^2/4$, $\delta=\varepsilon_1-\varepsilon_2-\Delta$, and $\Delta^2=(\varepsilon_1-\varepsilon_2)^2+4(\mathcal{V}_{12}-i\Gamma)^2$. One takes note that at either $\Gamma=0$ or $\varepsilon_1=\varepsilon_2$, Eq.~(\ref{Cpm}) leads to $ \mathcal{C}_{+-}=\mathcal{C}_{-+}=0$.

%\vspace{-0.2cm}

\begin{figure}[h!]
\begin{center}
\includegraphics[width=4.1cm]{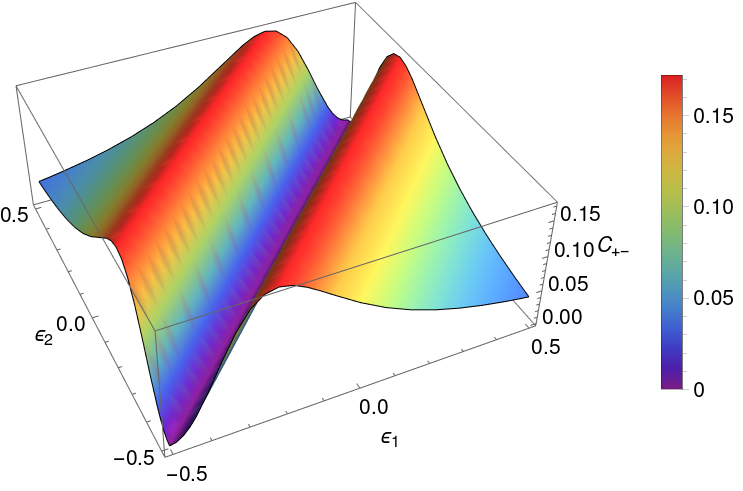}
\caption{Coherence factor $\mathcal{C}_{+-}=\mathcal{C}_{-+}$ for a DQD coupled in parallel at $\mathcal{V}_{12}=\Gamma=0.1$.}\label{fig_Cpm}
\end{center}
\end{figure}
\end{itemize}

\pagebreak

%%%%%%%%%%%%%%%%%%%%%%%%%%%%%%%%%%%%%%%%%%%%%%%%%%%%%%%%%%%%%%%%%%
%																 %
%																 %
%		X(w) for a DQD in series at T=0 						 %
%																 %
%																 %
%%%%%%%%%%%%%%%%%%%%%%%%%%%%%%%%%%%%%%%%%%%%%%%%%%%%%%%%%%%%%%%%%%

\section{Expression for $\mathcal{X}_c(\omega)$ at $T=0$}

For a DQD system, one has  $\mathcal{X}_c(\omega)=\sum_{s_1,s_2=\pm}  \mathcal{C}_{s_1s_2} \mathcal{X}_{s_1s_2}(\omega)$  with 
\begin{eqnarray}
\mathcal{X}_{s_1s_2}(\omega)=i\int_{-\infty}^\infty  \frac{d\varepsilon}{2\pi}  \Big[\mathcal{G}^<_{s_1s_2}(\varepsilon)\mathcal{G}^a_{s_2}(\varepsilon-\hbar\omega)+\mathcal{G}^r_{s_1}(\varepsilon+\hbar\omega)\mathcal{G}^<_{s_1s_2}(\varepsilon)\Big]~,
\end{eqnarray}
where $\mathcal{G}^r_{\pm}(\varepsilon)$ and $\mathcal{G}^a_{\pm}(\varepsilon)$ are the retarded and advanced Green functions in the dots given by $\mathcal{G}^r_{\pm}(\varepsilon)=(\varepsilon-\lambda_\pm)^{-1}$ and $\mathcal{G}^a_{\pm}(\varepsilon)=(\varepsilon-\lambda_\pm^*)^{-1}$, and $ \doubleunderline{\mathcal{G}}^<(\varepsilon)$ are the nonequilibrium Green functions in the dots which in the steady state are given by
\begin{eqnarray}
 \doubleunderline{\mathcal{G}}^<(\varepsilon)=\doubleunderline{\mathcal{G}}^r(\varepsilon)\doubleunderline U^{-1}\doubleunderline{\mathbb{\Sigma}}^<(\varepsilon)\,\doubleunderline U^{\,\text{ }}\doubleunderline{\mathcal{G}}^a(\varepsilon)~,
\end{eqnarray}
with  $\doubleunderline{\mathbb{\Sigma}}^<(\varepsilon)=i\sum_{\alpha=L,R}f_\alpha(\varepsilon)\doubleunderline\Gamma_{\,\alpha}$. The matrices $\doubleunderline{\mathcal{G}}^r(\varepsilon)$ and $\doubleunderline{\mathcal{G}}^a(\varepsilon)$ are diagonal while $\doubleunderline{\mathcal{G}}^<(\varepsilon)$ may be non-diagonal.

We set the constant $\hbar$ to 1 in the details of the calculation but we will restore this omitted factor $\hbar$ in the final expressions of the results.

\subsection{DQD in series}

For a DQD in series with symmetrically capacitive couplings $\alpha_1=\alpha_2=1$, one has $\Gamma_{L,11}=\Gamma_{R,22}\equiv \Gamma$ and $\Gamma_{L,22}=\Gamma_{R,11}=\Gamma_{\alpha,12}=0$, thus:
\begin{itemize}
 \item Coherence factors: $ \mathcal{C}_{++}= \mathcal{C}_{--}=1$ and $ \mathcal{C}_{+-}= \mathcal{C}_{-+}=0$.
\item $\doubleunderline\Gamma_{\,L}$ and $\doubleunderline\Gamma_{\,R}$ are diagonal matrices:
\begin{eqnarray}
\doubleunderline\Gamma_{\,L}=\left(
\begin{array}{ccc}
\Gamma& & 0\\
0& &0
\end{array}
\right)~,\quad
\doubleunderline\Gamma_{\,R}=\left(
\begin{array}{ccc}
0& & 0\\
0& &\Gamma
\end{array}
\right)\quad\Rightarrow\quad
\doubleunderline{\mathbb{\Sigma}}^<(\varepsilon)=
\left(
\begin{array}{ccc}
i\Gamma f_L(\varepsilon)& 0\\
0& i\Gamma f_R(\varepsilon)
\end{array}
\right)~.
\end{eqnarray}
\item Eigenvalues:
 \begin{eqnarray}\label{def_lambda}
\lambda_\pm= \frac{1}{2}\left(\varepsilon_1+\varepsilon_2-i\Gamma\pm \sqrt{(\varepsilon_1-\varepsilon_2)^2+4|\mathcal{V}_{12}|^2}\right)~.
\end{eqnarray}
\item Diagonalization matrix:
\begin{eqnarray}
\doubleunderline U=\frac{1}{\sqrt{|\mathcal{V}_{12}|^2+\delta^2/4}}\left(
\begin{array}{ccc}
\mathcal{V}^*_{12} & & \cfrac{\delta}{2}\\
-\cfrac{\delta}{2}& &\mathcal{V}_{12}
\end{array}
\right)\quad\Rightarrow\quad
\doubleunderline U^{-1}=
\frac{1}{\sqrt{|\mathcal{V}_{12}|^2+\delta^2/4}}\left(
\begin{array}{ccc}
\mathcal{V}_{12} & & -\cfrac{\delta}{2}\\
\cfrac{\delta}{2}& &\mathcal{V}^*_{12}
\end{array}
\right)~,
\end{eqnarray}
with $\delta=\varepsilon_1-\varepsilon_2-\Delta$, $\Delta^2=(\varepsilon_1-\varepsilon_2)^2+4|\mathcal{V}_{12}|^2$, and $\mathcal{V}^*_{21}=\mathcal{V}_{12}$.
\end{itemize}

It leads to
\begin{eqnarray}
 \doubleunderline{\mathcal{G}}^<(\varepsilon)&=&
 \frac{i\Gamma}{|\mathcal{V}_{12}|^2+\delta^2/4}
\left(
\begin{array}{ccc}
\mathcal{G}^r_{+}(\varepsilon)&  0\\
0& \mathcal{G}^r_{-}(\varepsilon)
\end{array}
\right)
\left(
\begin{array}{ccc}
\mathcal{V}_{12} & & -\frac{\delta}{2}\\
\frac{\delta}{2}& &\mathcal{V}^*_{12}
\end{array}
\right)
\left(
\begin{array}{ccc}
f_L(\varepsilon)& 0\\
0& f_R(\varepsilon)
\end{array}
\right)
\left(
\begin{array}{ccc}
\mathcal{V}^*_{12} & & \frac{\delta}{2}\\
-\frac{\delta}{2}& &\mathcal{V}_{12}
\end{array}
\right)
\left(
\begin{array}{ccc}
\mathcal{G}^a_{+}(\varepsilon)& 0\\
0& \mathcal{G}^a_{-}(\varepsilon)
\end{array}
\right)\nonumber\\
&=& \frac{i\Gamma}{|\mathcal{V}_{12}|^2+\delta^2/4}
\left(
\begin{array}{ccc}
\mathcal{G}^r_{+}(\varepsilon)&  0\\
0& \mathcal{G}^r_{-}(\varepsilon)
\end{array}
\right)
\left(
\begin{array}{ccc}
\mathcal{V}_{12} & & -\frac{\delta}{2}\\
\frac{\delta}{2}& &\mathcal{V}^*_{12}
\end{array}
\right)
\left(
\begin{array}{ccc}
f_L(\varepsilon)& 0\\
0& f_R(\varepsilon)
\end{array}
\right)
\left(
\begin{array}{ccc}
\mathcal{V}^*_{12}\mathcal{G}^a_{+}(\varepsilon) & & \frac{\delta}{2}\mathcal{G}^a_{-}(\varepsilon)\\
-\frac{\delta}{2}\mathcal{G}^a_{+}(\varepsilon)& &\mathcal{V}_{12}\mathcal{G}^a_{-}(\varepsilon)
\end{array}
\right)\nonumber\\
&=&\frac{i\Gamma}{|\mathcal{V}_{12}|^2+\delta^2/4}
\left(
\begin{array}{ccc}
\mathcal{G}^r_{+}(\varepsilon)&  0\\
0& \mathcal{G}^r_{-}(\varepsilon)
\end{array}
\right)
\left(
\begin{array}{ccc}
\mathcal{V}_{12} & & -\frac{\delta}{2}\\
\frac{\delta}{2}& &\mathcal{V}^*_{12}
\end{array}
\right)
\left(
\begin{array}{ccc}
\mathcal{V}^*_{12}f_L(\varepsilon)\mathcal{G}^a_{+}(\varepsilon) & & \frac{\delta}{2}f_L(\varepsilon)\mathcal{G}^a_{-}(\varepsilon)\\
-\frac{\delta}{2}f_R(\varepsilon)\mathcal{G}^a_{+}(\varepsilon)& &\mathcal{V}_{12}f_R(\varepsilon)\mathcal{G}^a_{-}(\varepsilon)
\end{array}
\right)\nonumber\\
&=&\frac{i\Gamma}{|\mathcal{V}_{12}|^2+\delta^2/4}
\left(
\begin{array}{ccc}
\mathcal{G}^r_{+}(\varepsilon)&  0\\
0& \mathcal{G}^r_{-}(\varepsilon)
\end{array}
\right)
\left(
\begin{array}{ccc}
[|\mathcal{V}_{12}|^2f_L(\varepsilon)+\frac{\delta^2}{4}f_R(\varepsilon)]\mathcal{G}^a_{+}(\varepsilon) & & \frac{\mathcal{V}_{12}\delta}{2}[f_L(\varepsilon)-f_R(\varepsilon)]\mathcal{G}^a_{-}(\varepsilon)\\
\frac{\mathcal{V}^*_{12}\delta}{2}[f_L(\varepsilon)-f_R(\varepsilon)]\mathcal{G}^a_{+}(\varepsilon)& &[|\mathcal{V}_{12}|^2f_R(\varepsilon)+\frac{\delta^2}{4}f_L(\varepsilon)]\mathcal{G}^a_{-}(\varepsilon)
\end{array}
\right)\nonumber\\
&=&\frac{i\Gamma}{|\mathcal{V}_{12}|^2+\delta^2/4}
\left(
\begin{array}{ccc}
[|\mathcal{V}_{12}|^2f_L(\varepsilon)+\frac{\delta^2}{4}f_R(\varepsilon)]|\mathcal{G}^r_{+}(\varepsilon)|^2 & & \frac{\mathcal{V}_{12}\delta}{2}[f_L(\varepsilon)-f_R(\varepsilon)]\mathcal{G}^r_{+}(\varepsilon)\mathcal{G}^a_{-}(\varepsilon)\\
\frac{\mathcal{V}^*_{12}\delta}{2}[f_L(\varepsilon)-f_R(\varepsilon)]\mathcal{G}^r_{-}(\varepsilon)\mathcal{G}^a_{+}(\varepsilon)& &[|\mathcal{V}_{12}|^2f_R(\varepsilon)+\frac{\delta^2}{4}f_L(\varepsilon)]|\mathcal{G}^r_{-}(\varepsilon)|^2
\end{array}
\right)~.
\end{eqnarray}

By using this result, one gets at $T=0$
\begin{eqnarray}
 &&\mathcal{X}_{++}(\omega)=-\frac{\Gamma}{|\mathcal{V}_{12}|^2+\delta^2/4}\int_{-\infty}^\infty  \frac{d\varepsilon}{2\pi}  [|\mathcal{V}_{12}|^2f_L(\varepsilon)+\delta^2f_R(\varepsilon)/4]|\mathcal{G}^r_{+}(\varepsilon)|^2\Big[\mathcal{G}^a_{+}(\varepsilon-\omega)+\mathcal{G}^r_{+}(\varepsilon+\omega)\Big]\nonumber\\
 &=&-\frac{\Gamma}{|\mathcal{V}_{12}|^2+\delta^2/4}\left(|\mathcal{V}_{12}|^2\int_{-\infty}^{\mu_L}  \frac{d\varepsilon}{2\pi}  \mathcal{G}^r_{+}(\varepsilon)|^2\Big[\mathcal{G}^a_{+}(\varepsilon-\omega)+\mathcal{G}^r_{+}(\varepsilon+\omega)\Big]
 +\frac{\delta^2}{4}\int_{-\infty}^{\mu_R}  \frac{d\varepsilon}{2\pi}  \mathcal{G}^r_{+}(\varepsilon)|^2\Big[\mathcal{G}^a_{+}(\varepsilon-\omega)+\mathcal{G}^r_{+}(\varepsilon+\omega)\Big]\right)~,\nonumber\\
\end{eqnarray}
and
\begin{eqnarray}
 &&\mathcal{X}_{--}(\omega)=-\frac{\Gamma}{|\mathcal{V}_{12}|^2+\delta^2/4}\int_{-\infty}^\infty  \frac{d\varepsilon}{2\pi}  [|\mathcal{V}_{12}|^2f_R(\varepsilon)+\delta^2f_L(\varepsilon)/4]|\mathcal{G}^r_{-}(\varepsilon)|^2\Big[\mathcal{G}^a_{-}(\varepsilon-\omega)+\mathcal{G}^r_{-}(\varepsilon+\omega)\Big]\nonumber\\
 &=&-\frac{\Gamma}{|\mathcal{V}_{12}|^2+\delta^2/4}\left(|\mathcal{V}_{12}|^2\int_{-\infty}^{\mu_R}  \frac{d\varepsilon}{2\pi}  \mathcal{G}^r_{-}(\varepsilon)|^2\Big[\mathcal{G}^a_{-}(\varepsilon-\omega)+\mathcal{G}^r_{-}(\varepsilon+\omega)\Big]
 +\frac{\delta^2}{4}\int_{-\infty}^{\mu_L}  \frac{d\varepsilon}{2\pi}  \mathcal{G}^r_{-}(\varepsilon)|^2\Big[\mathcal{G}^a_{-}(\varepsilon-\omega)+\mathcal{G}^r_{-}(\varepsilon+\omega)\Big]\right)~.\nonumber\\
\end{eqnarray}

One has to calculate the following integral
\begin{eqnarray}
I_{\alpha,\pm}(\omega)=\int_{-\infty}^{\mu_\alpha}  \frac{d\varepsilon}{2\pi}  \mathcal{G}^r_{\pm}(\varepsilon)|^2\Big[\mathcal{G}^a_{\pm}(\varepsilon-\omega)+\mathcal{G}^r_{\pm}(\varepsilon+\omega)\Big]~.
\end{eqnarray}

From Eq.~(\ref{def_lambda}), one can write
\begin{eqnarray}
 |\mathcal{G}^r_{\pm}(\varepsilon)|^2=|(\varepsilon-\lambda_\pm)^{-1}|^2=\frac{1}{\left(\varepsilon- \text{Re}\{\lambda_\pm\}\right)^2+\Gamma^2/4}~,
\end{eqnarray}
and
\begin{eqnarray}
 \mathcal{G}^a_{\pm}(\varepsilon-\omega)+\mathcal{G}^r_{\pm}(\varepsilon+\omega)=\frac{1}{\varepsilon-\omega-\text{Re}\{\lambda_\pm\}-i\Gamma/2}+\frac{1}{\varepsilon+\omega-\text{Re}\{\lambda_\pm\}+i\Gamma/2}~.
\end{eqnarray}

Thus
\begin{eqnarray}
 I_{\alpha,\pm}(\omega)=\int_{-\infty}^{\mu_\alpha} 
 \frac{d\varepsilon/(2\pi)}{\left(\varepsilon- \text{Re}\{\lambda_\pm\}\right)^2+\Gamma^2/4}
 \Big(\frac{1}{\varepsilon-\omega-\text{Re}\{\lambda_\pm\}-i\Gamma/2}+\frac{1}{\varepsilon+\omega-\text{Re}\{\lambda_\pm\}+i\Gamma/2}\Big)~.
\end{eqnarray}

By performing this integral explicitly, one gets
\begin{eqnarray}
I_{\alpha,\pm}(\omega)&=&\frac{1}{4\pi\omega(\Gamma-i\omega)}\Bigg[
 2\arctan\left(\frac{\Gamma/2}{\text{Re}\{\lambda_\pm\}-\mu_\alpha+\omega}\right)-2\arctan\left(\frac{\Gamma/2}{\text{Re}\{\lambda_\pm\}-\mu_\alpha-\omega}\right)\nonumber\\
 &&+i\ln\left(\frac{\left(\left(\mu_\alpha- \text{Re}\{\lambda_\pm\}\right)^2+\Gamma^2/4\right)^2}{\left(\left(\mu_\alpha-\omega- \text{Re}\{\lambda_\pm\}\right)^2+\Gamma^2/4\right)\left(\left(\mu_\alpha+\omega- \text{Re}\{\lambda_\pm\}\right)^2+\Gamma^2/4\right)}\right)
 \Bigg]~,
\end{eqnarray}
which can be written as
\begin{eqnarray}
I_{\alpha,\pm}(\omega)=-\frac{1}{2\pi\omega(\omega+i\Gamma)}
\ln\left(\frac{\left(\mu_\alpha- \text{Re}\{\lambda_\pm\}\right)^2+\Gamma^2/4}{\left(\mu_\alpha-\omega- \text{Re}\{\lambda_\pm\}-i\Gamma/2\right)\left(\mu_\alpha+\omega- \text{Re}\{\lambda_\pm\}+i\Gamma/2\right)}\right)~,
\end{eqnarray}
or as well
\begin{eqnarray}
I_{\alpha,\pm}(\omega)&=&-\frac{1}{2\pi\omega(\omega+i\Gamma)}
\ln\left(\frac{|\mu_\alpha-\lambda_\pm|^2}{\left(\mu_\alpha-\omega-\lambda_\pm^*\right)\left(\mu_\alpha+\omega-\lambda_\pm\right)}\right)
=\frac{1}{2\pi\omega(\omega+i\Gamma)}\ln\left((1-\omega \mathcal{G}^a_{\pm}(\mu_\alpha))(1+\omega \mathcal{G}^r_{\pm}(\mu_\alpha))\right)\nonumber\\
&=&\frac{1}{2\pi\omega(\omega+i\Gamma)}\ln\left(1-\omega\underbrace{ (\mathcal{G}^a_{\pm}(\mu_\alpha)- \mathcal{G}^r_{\pm}(\mu_\alpha))}_{=-2i\text{Im}\{ \mathcal{G}^r_{\pm}(\mu_\alpha)\}}-\omega^2| \mathcal{G}^r_{\pm}(\mu_\alpha)|^2\right)~.
\end{eqnarray}
It can be written equivalently in the form
\begin{eqnarray}
I_{\alpha,\pm}(\omega)=
\frac{1}{2\pi\omega(\omega+i\Gamma)}\ln\left(1-\frac{2\pi\omega(\omega+i\Gamma)}{\Gamma}A_\pm(\mu_\alpha)\right)~.
\end{eqnarray}

since one has  $A_\pm(\varepsilon)=-\text{Im}\{\mathcal{G}^r_{\pm}(\varepsilon)\}/\pi$ and for a DQD in series
\begin{eqnarray}
&& \mathcal{G}^r_{\pm}(\mu_\alpha)=\frac{1}{\mu_\alpha-\lambda_\pm}=\frac{1}{\mu_\alpha-\text{Re}\{\lambda_\pm\}+i\Gamma/2}
 =(\mu_\alpha-\text{Re}\{\lambda_\pm\}-i\Gamma/2)|\mathcal{G}^r_{\pm}(\mu_\alpha)|^2\nonumber\\
 &&\Rightarrow \text{Im}\{\mathcal{G}^r_{\pm}(\mu_\alpha)\}=-\frac{\Gamma}{2}|\mathcal{G}^r_{\pm}(\mu_\alpha)|^2
\Rightarrow 2\text{Im}\{\mathcal{G}^r_{\pm}(\mu_\alpha)\}+\Gamma|\mathcal{G}^r_{\pm}(\mu_\alpha)|^2=0~.
\end{eqnarray}

Finally, one gets $\mathcal{X}_c(\omega)=\mathcal{X}_{++}(\omega)+\mathcal{X}_{--}(\omega)$ with
\begin{eqnarray}
 \mathcal{X}_{++}(\omega)&=&-\frac{\Gamma}{2\pi\omega(\omega+i\Gamma)(|\mathcal{V}_{12}|^2+\delta^2/4)}\left(|\mathcal{V}_{12}|^2
\ln\left(1-\frac{2\pi\omega(\omega+i\Gamma)}{\Gamma}A_+(\mu_L)\right)
 +\frac{\delta^2}{4}
\ln\left(1-\frac{2\pi\omega(\omega+i\Gamma)}{\Gamma}A_+(\mu_R)\right)\right)~,\nonumber\\~\\
 \mathcal{X}_{--}(\omega)&=&-\frac{\Gamma}{2\pi\omega(\omega+i\Gamma)(|\mathcal{V}_{12}|^2+\delta^2/4)}\left(|\mathcal{V}_{12}|^2
\ln\left(1-\frac{2\pi\omega(\omega+i\Gamma)}{\Gamma}A_-(\mu_R)\right)
 +\frac{\delta^2}{4}
\ln\left(1-\frac{2\pi\omega(\omega+i\Gamma)}{\Gamma}A_-(\mu_L)\right)\right)~.\nonumber\\
\end{eqnarray}

The expression for $\mathcal{X}_c(\omega)$ can be simplified in the following two special cases (we restore the omitted $\hbar$ factor):
\begin{itemize}
 \item For $\varepsilon_1=\varepsilon_2=\varepsilon_0$, one has $\delta=-2\mathcal{V}_{12}$, which leads to
\begin{eqnarray}\label{expression_Xc}
 \boxed{\mathcal{X}_c(\omega)=
-\frac{\Gamma}{2h\omega(\hbar\omega+i\Gamma)}\sum_\pm\sum_{\alpha=L,R}\ln\left(1-\frac{h\omega(\hbar\omega+i\Gamma)}{\Gamma}A_\pm(\mu_\alpha)\right)}
\end{eqnarray}
\item For $\mu_L=\mu_R=0$ (zero-voltage), one gets whatever $\varepsilon_1$ and $\varepsilon_2$ are:
\begin{eqnarray}\label{expression_Xc_V0}
 \boxed{\mathcal{X}_c(\omega)=
-\frac{\Gamma}{h\omega(\hbar\omega+i\Gamma)}\sum_\pm\ln\left(1-\frac{h\omega(\hbar\omega+i\Gamma)}{\Gamma}A_\pm(0)\right)}
\end{eqnarray}
\end{itemize}

\subsection{DQD in parallel}

For a DQD in parallel with symmetrically capacitive couplings: $\alpha_1=\alpha_2=1$, identical couplings between each dot and each lead: $\Gamma_{\alpha,ij}\equiv \Gamma$ for $\forall \alpha,i,j$, and assuming coupling $\mathcal{V}_{12}$ to be real which implies $\mathcal{V}_{12}=\mathcal{V}_{21}$, one has:
\begin{itemize}
 \item Coherence factors: $ \mathcal{C}_{++}= \mathcal{C}_{--}=1$ and $ \mathcal{C}_{+-}= \mathcal{C}_{-+}=0$ since
 one has from Eq.~(\ref{Cpm})
\begin{eqnarray}
 \mathcal{C}_{+-}=\mathcal{C}_{-+}=\frac{|(\mathcal{V}_{12}-i\Gamma)\delta|^2-\text{Re}\{(\mathcal{V}_{12}+i\Gamma)^2\delta^2\}}{2D^4}~,
\end{eqnarray}
where $D^2=|\mathcal{V}_{12}-i\Gamma|^2+|\delta|^2/4$, $\delta=\varepsilon_1-\varepsilon_2-\Delta$, and $\Delta^2=(\varepsilon_1-\varepsilon_2)^2+4(\mathcal{V}_{12}-i\Gamma)^2$.

For $\varepsilon_1=\varepsilon_2=\varepsilon_0$, it leads to $\Delta=2(\mathcal{V}_{12}-i\Gamma)$, $\delta=-\Delta$ and $D^2=2|\mathcal{V}_{12}-i\Gamma|^2$, thus $ \mathcal{C}_{+-}=\mathcal{C}_{-+}=0$.
\item $\doubleunderline\Gamma_{\,L}$ and $\doubleunderline\Gamma_{\,R}$ are non-diagonal matrices equal to:
\begin{eqnarray}
\doubleunderline\Gamma_{\,L}=\left(
\begin{array}{ccc}
\Gamma & & \Gamma\\
\Gamma& & \Gamma 
\end{array}
\right)~,\quad
\doubleunderline\Gamma_{\,R}=\left(
\begin{array}{ccc}
\Gamma & & \Gamma\\
\Gamma& & \Gamma 
\end{array}
\right)\quad\Rightarrow\quad
\doubleunderline{\mathbb{\Sigma}}^<(\varepsilon)=i\sum_{\alpha=L,R} f_\alpha(\varepsilon)
\left(
\begin{array}{ccc}
\Gamma & & \Gamma\\
\Gamma& & \Gamma 
\end{array}
\right)~.
\end{eqnarray}
\item Eigenvalues:
\begin{eqnarray}
\lambda_\pm= \frac{1}{2}\left(\varepsilon_1+\varepsilon_2-2i\Gamma\pm\sqrt{(\varepsilon_1-\varepsilon_2)^2+4 (\mathcal{V}_{12}-i\Gamma)^2}\right)~.
\end{eqnarray}
%It leads for $\varepsilon_1=\varepsilon_2=\varepsilon_0$ to $\delta=-\Delta=-2(\mathcal{V}-i\Gamma)$, $D^2=2|\mathcal{V}-i\Gamma|^2$, thus
%$\lambda_\pm=\varepsilon_0-i\Gamma\pm (\mathcal{V}-i\Gamma)$.
%One has $\lambda_+=\varepsilon_0+\mathcal{V}-2i\Gamma$ and $\lambda_-=\varepsilon_0-\mathcal{V}$.
\item Diagonalization matrix:
\begin{eqnarray}
\doubleunderline U=\frac{1}{D}\left(
\begin{array}{ccc}
\mathcal{V}_{12}-i\Gamma& & \cfrac{\delta}{2}\\
-\cfrac{\delta}{2}& &\mathcal{V}_{12}-i\Gamma
\end{array}
\right)~,\quad
\doubleunderline U^{-1}=\frac{D}{(\mathcal{V}_{12}-i\Gamma)^2+\delta^2/4}\left(
\begin{array}{ccc}
\mathcal{V}_{12}-i\Gamma& & -\cfrac{\delta}{2}\\
\cfrac{\delta}{2}& &\mathcal{V}_{12}-i\Gamma
\end{array}
\right)~.
\end{eqnarray}
\end{itemize}

It leads to
\begin{eqnarray}
 \doubleunderline{\mathcal{G}}^<(\varepsilon)&=&
 \frac{i\sum_{\alpha=L,R}f_\alpha(\varepsilon)}{(\mathcal{V}_{12}-i\Gamma)^2+\delta^2/4}
\left(
\begin{array}{ccc}
\mathcal{G}^r_{+}(\varepsilon)&  0\\
0& \mathcal{G}^r_{-}(\varepsilon)
\end{array}
\right)
\left(
\begin{array}{ccc}
\mathcal{V}_{12}-i\Gamma& & -\frac{\delta}{2}\\
\frac{\delta}{2}& &\mathcal{V}_{12}-i\Gamma
\end{array}
\right)
\left(
\begin{array}{ccc}
1& 1\\
1& 1
\end{array}
\right)
\left(
\begin{array}{ccc}
\mathcal{V}_{12}-i\Gamma& & \frac{\delta}{2}\\
-\frac{\delta}{2}& &\mathcal{V}_{12}-i\Gamma
\end{array}
\right)
\left(
\begin{array}{ccc}
\mathcal{G}^a_{+}(\varepsilon)& 0\\
0& \mathcal{G}^a_{-}(\varepsilon)
\end{array}
\right)\nonumber\\
&=&
 \frac{i\sum_{\alpha=L,R}f_\alpha(\varepsilon)}{(\mathcal{V}_{12}-i\Gamma)^2+\delta^2/4}
\left(
\begin{array}{ccc}
\mathcal{G}^r_{+}(\varepsilon)&  0\\
0& \mathcal{G}^r_{-}(\varepsilon)
\end{array}
\right)
\left(
\begin{array}{ccc}
\mathcal{V}_{12}-i\Gamma& & -\frac{\delta}{2}\\
\frac{\delta}{2}& &\mathcal{V}_{12}-i\Gamma
\end{array}
\right)
\left(
\begin{array}{ccc}
1& 1\\
1& 1
\end{array}
\right)
\left(
\begin{array}{ccc}
(\mathcal{V}_{12}-i\Gamma)\mathcal{G}^a_{+}(\varepsilon)& & \frac{\delta}{2}\mathcal{G}^a_{-}(\varepsilon)\\
-\frac{\delta}{2}\mathcal{G}^a_{+}(\varepsilon)& &(\mathcal{V}_{12}-i\Gamma)\mathcal{G}^a_{-}(\varepsilon)
\end{array}
\right)\nonumber\\
&=&
 \frac{i\sum_{\alpha=L,R}f_\alpha(\varepsilon)}{(\mathcal{V}_{12}-i\Gamma)^2+\delta^2/4}
\left(
\begin{array}{ccc}
\mathcal{G}^r_{+}(\varepsilon)&  0\\
0& \mathcal{G}^r_{-}(\varepsilon)
\end{array}
\right)
\left(
\begin{array}{ccc}
\mathcal{V}_{12}-i\Gamma& & -\frac{\delta}{2}\\
\frac{\delta}{2}& &\mathcal{V}_{12}-i\Gamma
\end{array}
\right)
\left(
\begin{array}{ccc}
\left(\mathcal{V}_{12}-i\Gamma-\frac{\delta}{2}\right)\mathcal{G}^a_{+}(\varepsilon)& & \left(\mathcal{V}_{12}-i\Gamma+\frac{\delta}{2}\right)\mathcal{G}^a_{-}(\varepsilon)\\
\left(\mathcal{V}_{12}-i\Gamma-\frac{\delta}{2}\right)\mathcal{G}^a_{+}(\varepsilon)& & \left(\mathcal{V}_{12}-i\Gamma+\frac{\delta}{2}\right)\mathcal{G}^a_{-}(\varepsilon)
\end{array}
\right)\nonumber\\
&=&
 \frac{i\sum_{\alpha=L,R}f_\alpha(\varepsilon)}{(\mathcal{V}_{12}-i\Gamma)^2+\delta^2/4}
\left(
\begin{array}{ccc}
\mathcal{G}^r_{+}(\varepsilon)&  0\\
0& \mathcal{G}^r_{-}(\varepsilon)
\end{array}
\right)
\left(
\begin{array}{ccc}
\left(\mathcal{V}_{12}-i\Gamma-\frac{\delta}{2}\right)^2\mathcal{G}^a_{+}(\varepsilon)& & \left((\mathcal{V}_{12}-i\Gamma)^2-\frac{\delta^2}{4}\right)\mathcal{G}^a_{-}(\varepsilon)\\
\left((\mathcal{V}_{12}-i\Gamma)^2-\frac{\delta^2}{4}\right)\mathcal{G}^a_{+}(\varepsilon)& & \left(\mathcal{V}_{12}-i\Gamma+\frac{\delta}{2}\right)^2\mathcal{G}^a_{-}(\varepsilon)
\end{array}
\right)\nonumber\\
&=&
 \frac{i\sum_{\alpha=L,R}f_\alpha(\varepsilon)}{(\mathcal{V}_{12}-i\Gamma)^2+\delta^2/4}
\left(
\begin{array}{ccc}
\left(\mathcal{V}_{12}-i\Gamma-\frac{\delta}{2}\right)^2|\mathcal{G}^r_{+}(\varepsilon)|^2& & \left((\mathcal{V}_{12}-i\Gamma)^2-\frac{\delta^2}{4}\right)\mathcal{G}^r_{+}(\varepsilon)\mathcal{G}^a_{-}(\varepsilon)\\
\left((\mathcal{V}_{12}-i\Gamma)^2-\frac{\delta^2}{4}\right)\mathcal{G}^r_{-}(\varepsilon)\mathcal{G}^a_{+}(\varepsilon)& & \left(\mathcal{V}_{12}-i\Gamma+\frac{\delta}{2}\right)^2|\mathcal{G}^r_{-}(\varepsilon)|^2
\end{array}
\right)~,
\end{eqnarray}

which gives when $\varepsilon_1=\varepsilon_2=\varepsilon_0$
\begin{eqnarray}
 \doubleunderline{\mathcal{G}}^<(\varepsilon)=2i\Gamma\sum_{\alpha=L,R}f_\alpha(\varepsilon)\left(
\begin{array}{ccc}
|\mathcal{G}^r_{+}(\varepsilon)|^2& 0\\
0& 0
\end{array}
\right)~,
\end{eqnarray}
since one has $\delta=-\Delta=-2(\mathcal{V}_{12}-i\Gamma)$ in that case.

Using this result, one gets for $\mathcal{X}^r_c(\omega)$ at $T=0$
\begin{eqnarray}
\mathcal{X}^r_c(\omega)&=&-2\Gamma\sum_{\alpha=L,R}\int_{-\infty}^\infty \frac{d\varepsilon}{2\pi} f_\alpha(\varepsilon)|\mathcal{G}^r_{+}(\varepsilon)|^2\Big[\mathcal{G}^a_{+}(\varepsilon-\omega)+\mathcal{G}^r_{+}(\varepsilon+\omega)\Big]\nonumber\\
 &=&-2\Gamma\sum_{\alpha=L,R}\int_{-\infty}^{\mu_\alpha} \frac{d\varepsilon}{2\pi} |\mathcal{G}^r_{+}(\varepsilon)|^2\Big[\mathcal{G}^a_{+}(\varepsilon-\omega)+\mathcal{G}^r_{+}(\varepsilon+\omega)\Big]~.
\end{eqnarray}

By performing this integral explicitly, one gets
\begin{eqnarray}
 \mathcal{X}_c(\omega)&=&
-\frac{2\Gamma}{2\pi\omega(\omega+4i\Gamma)}\sum_{\alpha=L,R}\ln\left((1-\omega \mathcal{G}^a_{+}(\mu_\alpha))(1+\omega \mathcal{G}^r_{+}(\mu_\alpha))\right)\nonumber\\
&=&
-\frac{2\Gamma}{2\pi\omega(\omega+4i\Gamma)}\sum_{\alpha=L,R}\ln\left(1-\omega \underbrace{(\mathcal{G}^a_{+}(\mu_\alpha)- \mathcal{G}^r_{+}(\mu_\alpha))}_{=-2i\text{Im}\{ \mathcal{G}^r_+(\mu_\alpha)\}}-\omega^2|\mathcal{G}^r_{+}(\mu_\alpha)|^2\right)~,
\end{eqnarray}
which can be written equivalently in the form
\begin{eqnarray}\label{expression_Xc_parallel}
 \boxed{\mathcal{X}_c(\omega)=
-\frac{2\Gamma}{h\omega(\hbar\omega+4i\Gamma)}\sum_{\alpha=L,R}\ln\left(1-\frac{h\omega(\hbar\omega+4i\Gamma)}{4\Gamma}A_+(\mu_\alpha)\right)}
\end{eqnarray}
since one has for a DQD in parallel
\begin{eqnarray}
&& \mathcal{G}^r_{+}(\mu_\alpha)=\frac{1}{\mu_\alpha-\lambda_+}=\frac{1}{\mu_\alpha-\text{Re}\{\lambda_+\}+2i\Gamma}
 =(\mu_\alpha-\text{Re}\{\lambda_+\}-2i\Gamma)|\mathcal{G}^r_{+}(\mu_\alpha)|^2\nonumber\\
 &&\Rightarrow \text{Im}\{\mathcal{G}^r_{+}(\mu_\alpha)\}=-2\Gamma|\mathcal{G}^r_{+}(\mu_\alpha)|^2
\Rightarrow \text{Im}\{\mathcal{G}^r_{+}(\mu_\alpha)\}+2\Gamma|\mathcal{G}^r_{+}(\mu_\alpha)|^2=0~.
\end{eqnarray}

%%%%%%%%%%%%%%%%%%%%%%%%%%%%%%%%%%%%%%%%%%%%%%%%%%%%%%%%%%%%%%%%%%
%																 %
%																 %
%		RC-circuit						 %
%																 %
%																 %
%%%%%%%%%%%%%%%%%%%%%%%%%%%%%%%%%%%%%%%%%%%%%%%%%%%%%%%%%%%%%%%%%%

\section{Equivalent RC-circuit at $T=0$ and $\varepsilon_1=\varepsilon_2=\varepsilon_0$}

In this section, we calculate the capacitance and the resistance of the equivalent quantum RC-circuit which can be used to account for the behavior of the DQD system in the low-frequency regime\cite{Buttiker1993sm}.

%The results are
% \begin{itemize}
%  \item {\bf For DQD in series}\\
%  $C=(e^2/2)\sum_\pm\sum_\alpha A_\pm(\mu_\alpha)$ and $R=(h/e^2)\sum_\pm\sum_{\alpha}A^2_{\pm}(\mu_\alpha)/(\sum_\pm\sum_{\alpha}A_{\pm}(\mu_\alpha))^2$, where $A_\pm(\varepsilon)=-\text{Im}\{\mathcal{G}^r_\pm(\varepsilon)\}/\pi$ is the spectral function. One has $R\in[h/4e^2,h/e^2]$.
%  \item {\bf For DQD in parallel}\\
%  $C=(e^2/2)\sum_\alpha A_+(\mu_\alpha)$ and $R=(h/e^2)\sum_{\alpha}A^2_{+}(\mu_\alpha)/(\sum_{\alpha}A_{+}(\mu_\alpha))^2$. One has in that case $R\in[h/2e^2,h/e^2]$.
% \end{itemize}

\subsection{DQD in series}

%%%%%%%%%%%%%%%%%%%%%%%%%%%%%%%%%%%%%%%%%%%%%%%%%%%%%%%%%%%%%%%%%%
%																 %
%																 %
%		Cq	in series   										 %
%																 %
%																 %
%%%%%%%%%%%%%%%%%%%%%%%%%%%%%%%%%%%%%%%%%%%%%%%%%%%%%%%%%%%%%%%%%%

\subsubsection{$C$ for a DQD in series}

One has $C=e^2 \mathcal{X}_c(\omega=0)$. One gets from Eq.~(\ref{expression_Xc})
\begin{eqnarray}
 C=-\lim_{\omega\rightarrow 0}\frac{e^2\Gamma}{2h\omega(\hbar\omega+i\Gamma)}\sum_\pm\sum_{\alpha=L,R}\ln\left(1-\frac{h\omega(\hbar\omega+i\Gamma)}{\Gamma}A_\pm(\mu_\alpha)\right)
 \Rightarrow \boxed{C=\frac{e^2}{2}\sum_\pm\sum_{\alpha=L,R}A_\pm(\mu_\alpha)}
\end{eqnarray}
using $\ln(1+x)\approx x$.

%One underlines that $ \mathcal{X}_c(\omega=0)=\sum_{\alpha=L,R}A(\mu_\alpha)/2$, where $A(\mu_\alpha)=\sum_\pm A_\pm(\mu_\alpha)$.

%%%%%%%%%%%%%%%%%%%%%%%%%%%%%%%%%%%%%%%%%%%%%%%%%%%%%%%%%%%%%%%%%%
%																 %
%																 %
%		Rq	in series										 %
%																 %
%																 %
%%%%%%%%%%%%%%%%%%%%%%%%%%%%%%%%%%%%%%%%%%%%%%%%%%%%%%%%%%%%%%%%%%

\subsubsection{$R$ for a DQD in series}

One has
\begin{eqnarray}
 R=\cfrac{e^2}{C^2}\lim_{\omega\rightarrow 0}\cfrac{\text{Im}\{\mathcal{X}_c(\omega)\}}{\omega }~.
\end{eqnarray}
By using $\ln(1+x)\approx x-x^2/2$, one gets from Eq.~(\ref{expression_Xc})
\begin{eqnarray}
 \mathcal{X}_c(\omega)
=-\frac{\Gamma}{2h\omega(\hbar\omega+i\Gamma)}\sum_\pm\sum_{\alpha=L,R}\left(-\frac{h\omega(\hbar\omega+i\Gamma)}{\Gamma}A_\pm(\mu_\alpha)-\frac{h^2\omega^2(i\Gamma)^2}{2\Gamma^2}A^2_\pm(\mu_\alpha)\right)~.
\end{eqnarray}
It leads to
\begin{eqnarray}
 \mathcal{X}_c(\omega)
=-\frac{\Gamma(\hbar\omega-i\Gamma)}{2h\omega(\hbar\omega^2+\Gamma^2)}\sum_\pm\sum_{\alpha=L,R}\left(-\frac{h\omega(\hbar\omega+i\Gamma)}{\Gamma}A_\pm(\mu_\alpha)+\frac{h^2\omega^2}{2} A^2_\pm(\mu_\alpha)\right)~.
\end{eqnarray}
The imaginary part of $ \mathcal{X}_c(\omega)$ is then given by
\begin{eqnarray}
\text{Im}\{\mathcal{X}_c(\omega)\}=\frac{\Gamma^2}{4h\omega(\hbar^2\omega^2+\Gamma^2)}\sum_\pm\sum_{\alpha=L,R}\left(h^2\omega^2 A^2_\pm(\mu_\alpha)\right)
=\frac{h\omega\Gamma^2}{4(\hbar^2\omega^2+\Gamma^2)}\sum_\pm\sum_{\alpha=L,R}A^2_\pm(\mu_\alpha)~.
\end{eqnarray}
Thus
\begin{eqnarray}
 \lim_{\omega\rightarrow 0}\cfrac{\text{Im}\{\mathcal{X}_c(\omega)\}}{\omega }
 &=&\frac{h}{4}\sum_\pm\sum_{\alpha=L,R}A^2_\pm(\mu_\alpha)~.
\end{eqnarray}
By using $C=(e^2/2)\sum_\pm\sum_{\alpha=L,R}A_\pm(\mu_\alpha)$, one finally gets
\begin{eqnarray}\label{Rq_exp}
\boxed{R=
\frac{h}{e^2}\cfrac{\sum_\pm\sum_{\alpha=L,R}A^2_\pm(\mu_\alpha)}{(\sum_\pm\sum_{\alpha=L,R}A_\pm(\mu_\alpha))^2}}
\end{eqnarray}

The Cauchy-Schwarz inequality $(\sum_{i=1}^n x_i^2)(\sum_{i=1}^n y_i^2)\ge(\sum_{i=1}^n x_iy_i)^2$ for $x_i=A_\pm(\mu_\alpha)$ and $y_i=1$  leads to
\begin{eqnarray}
&&\bigg(\sum_\pm\sum_{\alpha=L,R}A^2_\pm(\mu_\alpha)\bigg)\underbrace{\bigg(\sum_\pm\sum_{\alpha=L,R}1^2\bigg)}_{=4}
\ge (\sum_\pm\sum_{\alpha=L,R}A_\pm(\mu_\alpha)\times 1)^2\nonumber\\
&&\Rightarrow \cfrac{\sum_\pm\sum_{\alpha=L,R}A^2_\pm(\mu_\alpha)}{(\sum_\pm\sum_{\alpha=L,R}A_\pm(\mu_\alpha))^2} \ge \frac{1}{4} \Rightarrow R \ge \frac{h}{4e^2}~,
\end{eqnarray}
Moreover, in all generality one has $\sum_{i=1}^n x_i^4\le (\sum_{i=1}^n x_i^2)^2$, thus
\begin{eqnarray}
 &&\sum_\pm\sum_{\alpha=L,R}|\mathcal{G}^r_{\pm}(\mu_\alpha)|^4\le (\sum_\pm\sum_{\alpha=L,R}|\mathcal{G}^r_{\pm}(\mu_\alpha)|^2)^2\nonumber\\
 && \Rightarrow   \frac{\sum_\pm\sum_{\alpha=L,R}A^2_\pm(\mu_\alpha)}{(\sum_\pm\sum_{\alpha=L,R}A_\pm(\mu_\alpha))^2}=
 \cfrac{\sum_\pm\sum_{\alpha=L,R}|\mathcal{G}^r_{\pm}(\mu_\alpha)|^4}{(\sum_\pm\sum_{\alpha=L,R}|\mathcal{G}^r_{\pm}(\mu_\alpha)|^2)^2}
\le 1
 \Rightarrow R\le  \frac{h}{e^2}~,
\end{eqnarray}
since one has for a DQD in series: $A_\pm(\varepsilon)=-\text{Im}\{\mathcal{G}^r_{\pm}(\varepsilon)\}/\pi=(\Gamma/2\pi)|\mathcal{G}^r_{\pm}(\varepsilon)|^2$.

In conclusion, one has $R\in[h/4e^2,h/e^2]$ for a DQD in series at $T=0$ and $\varepsilon_1=\varepsilon_2$.

%%%%%%%%%%%%%%%%%%%%%%%%%%%%%%%%%%%%%%%%%%%%%%%%%%%%%%%%%%%%%%%%%%
%																 %
%																 %
%		DQD in parallel											 %
%																 %
%																 %
%%%%%%%%%%%%%%%%%%%%%%%%%%%%%%%%%%%%%%%%%%%%%%%%%%%%%%%%%%%%%%%%%%

\subsection{DQD in parallel}

%%%%%%%%%%%%%%%%%%%%%%%%%%%%%%%%%%%%%%%%%%%%%%%%%%%%%%%%%%%%%%%%%%
%																 %
%																 %
%		Cq											 %
%																 %
%																 %
%%%%%%%%%%%%%%%%%%%%%%%%%%%%%%%%%%%%%%%%%%%%%%%%%%%%%%%%%%%%%%%%%%

\subsubsection{$C$ for a DQD in parallel}

One has $C=e^2 \mathcal{X}_c(\omega=0)$. One gets from Eq.~(\ref{expression_Xc_parallel})
\begin{eqnarray}
 C=-\lim_{\omega\rightarrow 0}\frac{2e^2\Gamma}{h\omega(\hbar\omega+4i\Gamma)}\sum_{\alpha=L,R}\ln\left(1-\frac{h\omega(\hbar\omega+4i\Gamma)}{4\Gamma}A_+(\mu_\alpha)\right)
 \Rightarrow \boxed{C=\frac{e^2}{2}\sum_{\alpha=L,R}A_+(\mu_\alpha)}
\end{eqnarray}
using $\ln(1+x)\approx x$.

%%%%%%%%%%%%%%%%%%%%%%%%%%%%%%%%%%%%%%%%%%%%%%%%%%%%%%%%%%%%%%%%%%
%																 %
%																 %
%		Rq											 %
%																 %
%																 %
%%%%%%%%%%%%%%%%%%%%%%%%%%%%%%%%%%%%%%%%%%%%%%%%%%%%%%%%%%%%%%%%%%

\subsubsection{$R$ for a DQD in parallel}

One has
\begin{eqnarray}
 R=\cfrac{e^2}{C^2}\lim_{\omega\rightarrow 0}\cfrac{\text{Im}\{\mathcal{X}_c(\omega)\}}{\omega }~.
\end{eqnarray}
By using $\ln(1+x)\approx x-x^2/2$, one gets from Eq.~(\ref{expression_Xc})
\begin{eqnarray}
 \mathcal{X}_c(\omega)
=-\frac{2\Gamma}{h\omega(\hbar\omega+4i\Gamma)}\sum_{\alpha=L,R}\left(-\frac{h\omega(\hbar\omega+4i\Gamma)}{\Gamma}A_+(\mu_\alpha)-\frac{h^2\omega^2(4i\Gamma)^2}{32\Gamma^2}A^2_+(\mu_\alpha)\right)~.
\end{eqnarray}
It leads to
\begin{eqnarray}
 \mathcal{X}_c(\omega)
=-\frac{2\Gamma(\hbar\omega-4i\Gamma)}{h\omega(\hbar\omega^2+16\Gamma^2)}\sum_{\alpha=L,R}\left(-\frac{h\omega(\hbar\omega+4i\Gamma)}{\Gamma}A_+(\mu_\alpha)+\frac{h^2\omega^2}{2} A^2_+(\mu_\alpha)\right)~.
\end{eqnarray}
The imaginary part of $ \mathcal{X}_c(\omega)$ is then given by
\begin{eqnarray}
\text{Im}\{\mathcal{X}_c(\omega)\}=\frac{4\Gamma^2}{h\omega(\hbar^2\omega^2+16\Gamma^2)}\sum_{\alpha=L,R}\left(h^2\omega^2 A^2_+(\mu_\alpha)\right)
=\frac{4h\omega\Gamma^2}{(\hbar^2\omega^2+16\Gamma^2)}\sum_{\alpha=L,R}A^2_+(\mu_\alpha)~.
\end{eqnarray}
Thus
\begin{eqnarray}
 \lim_{\omega\rightarrow 0}\cfrac{\text{Im}\{\mathcal{X}_c(\omega)\}}{\omega }
 &=&\frac{h}{4}\sum_{\alpha=L,R}A^2_+(\mu_\alpha)~.
\end{eqnarray}
By using $C=(e^2/2)\sum_{\alpha=L,R}A(\mu_\alpha)$, one finally gets
\begin{eqnarray}\label{Rq_exp}
\boxed{R=
\frac{h}{e^2}\cfrac{\sum_{\alpha=L,R}A^2_+(\mu_\alpha)}{(\sum_{\alpha=L,R}A_+(\mu_\alpha))^2}}
\end{eqnarray}

The Cauchy-Schwarz inequality $(\sum_{i=1}^n x_i^2)(\sum_{i=1}^n y_i^2)\ge(\sum_{i=1}^n x_iy_i)^2$ for $x_i=A_+(\mu_\alpha)$ and $y_i=1$ leads to
\begin{eqnarray}
&&\bigg(\sum_{\alpha=L,R}A^2_+(\mu_\alpha)\bigg)\underbrace{\bigg(\sum_{\alpha=L,R}(1)^2\bigg)}_{=2}
\ge (\sum_{\alpha=L,R}A_+(\mu_\alpha)\times 1)^2\nonumber\\
&&\Rightarrow \cfrac{\sum_{\alpha=L,R}A^2_+(\mu_\alpha)}{(\sum_{\alpha=L,R}A_+(\mu_\alpha))^2} \ge \frac{1}{2} \Rightarrow R \ge \frac{h}{2e^2}~,
\end{eqnarray}
Moreover, one has in all generality $\sum_{i=1}^n x_i^4\le (\sum_{i=1}^n x_i^2)^2$, thus
\begin{eqnarray}
 &&\sum_{\alpha=L,R}|\mathcal{G}^r_{+}(\mu_\alpha)|^4\le (\sum_{\alpha=L,R}|\mathcal{G}^r_{+}(\mu_\alpha)|^2)^2\nonumber\\
 && \Rightarrow  \frac{\sum_\pm\sum_{\alpha=L,R}A^2_+(\mu_\alpha)}{(\sum_\pm\sum_{\alpha=L,R}A_+(\mu_\alpha))^2}=
 \cfrac{\sum_{\alpha=L,R}|\mathcal{G}^r_{+}(\mu_\alpha)|^4}{(\sum_{\alpha=L,R}|\mathcal{G}^r_{+}(\mu_\alpha)|^2)^2} \le 1
 \Rightarrow R\le  \frac{h}{e^2}~,
\end{eqnarray}
since one has for a DQD in parallel: $A_+(\varepsilon)=-\text{Im}\{\mathcal{G}^r_{+}(\varepsilon)\}/\pi=(2\Gamma/\pi)|\mathcal{G}^r_{+}(\varepsilon)|^2$.

In conclusion, one has $R\in[h/2e^2,h/e^2]$ for a DQD in parallel at $T=0$ and $\varepsilon_1=\varepsilon_2$.

%%%%%%%%%%%%%%%%%%%%%%%%%%%%%%%%%%%%%%%%%%%%%%%%%%%%%%%%%%%%%%%%%%
%																 %
%																 %
%		Singulet+Triplet states						 %
%																 %
%																 %
%%%%%%%%%%%%%%%%%%%%%%%%%%%%%%%%%%%%%%%%%%%%%%%%%%%%%%%%%%%%%%%%%%

\section{Phase response $\phi(\omega)$ of the resonator}

By definition, the phase response of the resonator is equal~to
\begin{eqnarray}\label{phase_def}
 \phi(\omega)=\arctan\left(\frac{\text{Im}\{Z(\omega)\}}{\text{Re}\{Z(\omega)\}}\right)~,
\end{eqnarray}
where $Z(\omega)$ is the impedance defined as $Z(\omega)=dV_\text{gate}(\omega)/dI(\omega)$. Given that the dynamical charge susceptibility is $\mathcal{X}_c(\omega)=d N(\omega)/d(eV_\text{gate}(\omega))$\cite{Lavagna2020sm} and that the electrical current is defined as $I(t)=-dQ(t)/dt=-edN(t)/dt \Rightarrow I(\omega)=-ie\omega N(\omega)$, one gets
\begin{eqnarray}
 \mathcal{X}_c(\omega)=\frac{d N(\omega)}{d(eV_\text{gate}(\omega))}=\frac{1}{e}\frac{d N(\omega)}{d I(\omega)}\frac{d I(\omega)}{d V_\text{gate}(\omega)}
 =-\frac{1}{ie^2\omega Z(\omega)}=\frac{iZ^*(\omega)}{e^2\omega |Z(\omega)|^2}~.
\end{eqnarray}
Thus
\begin{eqnarray}
  \text{Re}\{Z(\omega)\}= e^2\omega|Z(\omega)|^2\text{Im}\{\mathcal{X}_c(\omega)\}\quad\text{and}\quad
    \text{Im}\{Z(\omega)\}=e^2\omega|Z(\omega)|^2\text{Re}\{\mathcal{X}_c(\omega)\}~.
\end{eqnarray}
By incorporating these latter expressions into Eq.~(\ref{phase_def}), one gets
\begin{eqnarray}
 \phi(\omega)=\arctan\left(\frac{\text{Re}\{\mathcal{X}_c(\omega)\}}{\text{Im}\{\mathcal{X}_c(\omega)\}}\right)~.
\end{eqnarray}

%%%%%%%%%%%%%%%%%%%%%%%%%%%%%%%%%%%%%%%%%%%%%%%%%%%%%%%%%%%%%%%%%%
%																 %
%																 %
%		Singulet+Triplet states						 %
%																 %
%																 %
%%%%%%%%%%%%%%%%%%%%%%%%%%%%%%%%%%%%%%%%%%%%%%%%%%%%%%%%%%%%%%%%%%

\section{Expression for $\mathcal{X}_c(\omega)$ in a DQD in series in the presence of additional triplet states}

In order to compare with experiments performed in spin qubits, one needs to take triplet states into account in addition to the singlet states. To do this, one includes an additional term to the expression for the dynamical charge susceptibility given in (\ref{expression_Xc_V0}) for a DQD in series at $T=0$ and when $\varepsilon_1=\varepsilon_2$, corresponding to the triplet state contribution, as follows
\begin{eqnarray}\label{Xc_triplet}
\mathcal{X}_c(\omega)=
-\frac{\Gamma}{2h\omega(\hbar\omega+i\Gamma)}\sum_{\alpha=L,R}\left(\sum_\pm\ln\left(1-\frac{h\omega(\hbar\omega+i\Gamma)}{\Gamma}A_\pm(\mu_\alpha)\right)
+3\ln\left(1-\frac{h\omega(\hbar\omega+i\Gamma)}{\Gamma}A_T(\mu_\alpha)\right)\right)~,
\end{eqnarray}
where $A_T(\varepsilon)=-\text{Im}\{\mathcal{G}_T^r(\varepsilon)\}/\pi$ is the spectral function associated with the Green function $\mathcal{G}_T^r(\varepsilon)=(\varepsilon-\lambda_T)^{-1}$ where $\lambda_T=(\varepsilon_2-\varepsilon_1-i\Gamma)/2$ is the energy of the triplet states. The factor 3 in front of the second term in the r.h.s. of Eq.~(\ref{Xc_triplet}) results from the fact that the number of states in the triplet states is equal to 3.

\bibliographystyle{apsrev4-1}
%\bibnote[]{} pour inserer une note dans le texte.

\end{document}